\documentclass[aps,prb,twocolumn,superscriptaddress,amsmath,amssymb]{revtex4}

\usepackage{amssymb}
\usepackage{graphicx}
\usepackage{tabularx}
\usepackage{bm}
\usepackage{epsfig}
\usepackage[ansinew]{inputenc}
\usepackage{float}

\begin{document}

\title{The zero-field-cooled exchange bias effect in phase-segregated         
\\* La$_{2-x}$A$_{x}$CoMnO$_{6-\delta}$ (A = Ba, Ca, Sr; x = 0, 0.5)}

\author{L. T. Coutrim}
\affiliation{Instituto de F\'{\i}sica, Universidade Federal de Goi\'{a}s, 74001-970, Goi\^{a}nia, GO, Brazil}
\affiliation{Laborat{\'o}rio Nacional de Luz S\'{\i}ncrotron, Centro Nacional de Pesquisa em Energia e Materiais, 13083-970, Campinas, SP, Brazil}

\author{D. Rigitano}
\affiliation{Instituto de F\'{\i}sica ``Gleb Wataghin", UNICAMP, 13083-859, Campinas, SP, Brazil}

\author{C. Macchiutti}
\affiliation{Centro Brasileiro de Pesquisas F\'{\i}sicas, 22290-180, Rio de Janeiro, RJ, Brazil}

\author{T. J. A. Mori}
\affiliation{Laborat{\'o}rio Nacional de Luz S\'{\i}ncrotron, Centro Nacional de Pesquisa em Energia e Materiais, 13083-970, Campinas, SP, Brazil}

\author{R. Lora-Serrano}
\affiliation{Universidade Federal de Uberl\^{a}ndia, Instituto de F\'{\i}sica, 38400-902, Uberl\^{a}ndia-MG, Brazil}

\author{E. Granado}
\affiliation{Instituto de F\'{\i}sica ``Gleb Wataghin", UNICAMP, 13083-859, Campinas, SP, Brazil}

\author{E. Sadrollahi}
\affiliation{Institut f{\"u}r Physik der Kondensierten Materie, Technische Universit{\"a}t Braunschweig, 38110 Braunschweig, Germany}

\author{F. J. Litterst}
\affiliation{Centro Brasileiro de Pesquisas F\'{\i}sicas, 22290-180, Rio de Janeiro, RJ, Brazil}
\affiliation{Institut f{\"u}r Physik der Kondensierten Materie, Technische Universit{\"a}t Braunschweig, 38110 Braunschweig, Germany}

\author{M. B. Fontes}
\affiliation{Centro Brasileiro de Pesquisas F\'{\i}sicas, 22290-180, Rio de Janeiro, RJ, Brazil}

\author{E. Baggio-Saitovitch}
\affiliation{Centro Brasileiro de Pesquisas F\'{\i}sicas, 22290-180, Rio de Janeiro, RJ, Brazil}

\author{E. M. Bittar}
\affiliation{Centro Brasileiro de Pesquisas F\'{\i}sicas, 22290-180, Rio de Janeiro, RJ, Brazil}

\author{L. Bufai\c{c}al}
\email{lbufaical@ufg.br}
\affiliation{Instituto de F\'{\i}sica, Universidade Federal de Goi\'{a}s, 74001-970, Goi\^{a}nia, GO, Brazil}

\date{\today}

\begin{abstract}
In the zero-field-cooled exchange bias (ZEB) effect the unidirectional magnetic anisotropy is set at low temperatures even when the system is cooled in the absence of external magnetic field. La$_{1.5}$Sr$_{0.5}$CoMnO$_{6}$ stands out as presenting the largest ZEB reported so far, while for La$_{1.5}$Ca$_{0.5}$CoMnO$_{6}$ the exchange bias field ($H_{EB}$) is one order of magnitude smaller. Here we show that La$_{1.5}$Ba$_{0.5}$CoMnO$_{6}$ also exhibits a pronounced shift of its magnetic hysteresis loop, with intermediate $H_{EB}$ value in respect to Ca- and Sr-doped samples. In order to figure out the microscopic mechanisms responsible for this phenomena, these compounds were investigated by means of synchrotron X-ray powder diffraction, Raman spectroscopy, muon spin rotation and relaxation, AC and DC magnetization, X-ray absorption spectroscopy (XAS) and X-ray magnetic circular dichroism (XMCD). The parent compound La$_{2}$CoMnO$_{6}$ was also studied for comparison, as a reference of a non-ZEB material.  Our results show that the Ba-, Ca- and Sr-doped samples present a small amount of phase segregation, and that the ZEB effect is strongly correlated to the system's structure. We also observed that mixed valence states Co$^{2+}$/Co$^{3+}$ and Mn$^{4+}$/Mn$^{3+}$ are already present at the La$_{2}$CoMnO$_{6}$ parent compound, and that Ba$^{2+}$/Ca$^{2+}$/Sr$^{2+}$ partial substitution at La$^{3+}$ site leads to a large increase of Co average valence, with a subtle augmentation of Mn formal valence. Estimates of the Co and Mn valences from the $L$-edge XAS indicate the presence of oxygen vacancies in all samples (0.05$\leq \delta \leq$0.1). Our XMCD results show a great decrease of Co moment for the doped compounds, and indicate that the shift of the hysteresis curves for these samples is related to uncompensated antiferromagnetic coupling between Co and Mn.

\end{abstract}

%\pacs{75.50.Lk, 75.30.Gw, 75.60.Jk, 75.47.Lx}

\maketitle

\section{Introduction}

The exchange bias (EB) effect was extensively investigated over the past 60 years due to its applicability in high-density magnetic recording, giant magnetoresistance and spin valve devices \cite{Dagotto}. It consists in a non-equilibrium phenomenon caused by the uncompensated coupling at the interface between different magnetic components \cite{Nogues}, thus being observed in magnetic heterostructures as ferromagnetic (FM)-antiferromagnetic (AFM), FM-spin glass (SG), FM-ferrimagnetic (FIM), AFM-FIM, FIM-SG-AFM etc. The EB effect is manifested by a shift of the magnetic hysteresis loops along the field axis, being conventionally observed after cooling the system in the presence of a static magnetic field ($H$) from above its magnetic transition temperature ($T$). For many decades, the cooling $H$ was understood as a necessary protocol to set the unidirectional anisotropy (UA) by breaking the symmetry of the interface moment, and any shifted hysteresis loop observed after zero-field-cooling (ZFC) the system was naturally attributed to experimental artifacts and/or minor loop effects \cite{Geshev}.

In 2007 J. Saha \textit{et al.} proposed a model predicting UA in a ZFC FM-AFM system, induced by the application of the first field during the magnetization as a function of field [$M(H)$] measurement (the so-called virgin curve) \cite{Saha}. After that, robust ZFC EB (ZEB) has been experimentally observed in different kinds of materials as polycrystalline oxides \cite{Giri,CoIr_PRB}, nanocomposites \cite{Maity} alloys \cite{Wang,Nayak} and films \cite{Prieto}. The presence of SG-like phases, concomitant to other conventional phases as FM/AFM, seems to be a common feature in many of these materials, and recently we have shown the significant influence of glassy magnetism on the ZEB effect observed in double-perovskite (DP) compounds \cite{ZEBmodel}.

Due to their intrinsic inhomogeneity, DP materials usually present structural and magnetic disorder \cite{Vasala,Serrate}, which may lead to competing magnetic interactions and frustration. These characteristics are key ingredients to the appearance of SG-like behavior, and is not surprising that a great majority of ZEB materials have perovskite structure \cite{Giri,CoIr_PRB,Maity,Huang,Xie,Murthy,CaCoMn_JMMM}. Among the ZEB materials discovered so far, La$_{1.5}$Sr$_{0.5}$CoMnO$_{6}$ (LSCMO) presents the largest shift of the $M(H)$ curve \cite{Murthy}. Interestingly, we showed that replacing Sr by Ca also gives rise to a ZEB material, but the smaller Ca ionic radius leads to a shift of the $M(H)$ loop that is one order of magnitude smaller than that of the Sr-based compound \cite{CaCoMn_JMMM}. This exemplifies the strong correlation between structural, electronic and magnetic properties in these DP systems. Although it is a common ground that the re-entrant SG (RSG) behavior is necessary to the observation of the ZEB effect in these materials \cite{ZEBmodel}, the microscopic mechanisms responsible for such different magnetic properties observed in similar compounds is not well understood. 

In order to get further insight to this question, we thoroughly investigated La$_{1.5}$A$_{0.5}$CoMnO$_{6}$ (A = Ba, Ca, Sr) polycrystalline samples by means of synchrotron X-ray powder diffraction (SXRD), Raman spectroscopy, muon spin rotation and relaxation ($\mu$SR), AC and DC magnetization, X-ray absorption spectroscopy (XAS) at Co- and Mn-$K$ and $L_{3,2}$ edges, as well as  X-ray magnetic circular dichroism (XMCD) at Co- and Mn-$L_{3,2}$ edges. We show that La$_{1.5}$Ba$_{0.5}$CoMnO$_{6}$ (LBCMO) exhibits a giant ZEB effect, although not as large as that observed for LSCMO. The parent compound La$_{2}$CoMnO$_{6}$ (LCMO), which is a non-RSG and non-ZEB material, was also investigated, for comparison. For the SXRD measurements, we carried anomalous scattering with energy $E$ = 6500 eV, in order to maximize the difference between Co and Mn scattering factors and investigate whether these ions are ordered along the lattice or not. Our results indicate disordering of Co and Mn in all samples, and Ba-, Ca- and Sr-doped samples present a small amount of phase segregation. 

Since the ZEB effect is observed at low $T$, we also performed SXRD at several $T$ with $E$ = 9000 eV, and the results indicate that the larger ZEB effect observed for LBCMO and LSCMO is related to their more symmetrical crystal structures. These results are corroborated by the Raman spectroscopy measurements. The XAS results indicate mixed valence states Co$^{2+}$/Co$^{3+}$ and Mn$^{4+}$/Mn$^{3+}$ in all samples, and also that Ba$^{2+}$/Ca$^{2+}$/Sr$^{2+}$ partial substitution at La$^{3+}$ site leads to a main increase of Co average valence, while the increase of Mn formal valence is very subtle. Oxygen vacancies (OV) are present in all samples, which may play an important role in their magnetic properties.

All samples present two FM transitions, attributed to Co$^{2+}$--O--Mn$^{4+}$ and Co$^{3+}$--O--Mn$^{3+}$ couplings. For the Ba-,Ca- and Sr-doped compounds we observed anomalies in the temperature dependence of magnetization [$M(T)$], which were confirmed by the $\mu$SR results to be related to the onset of a third magnetic transition to a spin glassy state, possibly caused by the Co$^{3+}$--O--Mn$^{4+}$ AFM coupling. Our XMCD results show that the shift of the hysteresis curves are related to uncompensated AFM coupling between Co and Mn. The great decrease of Co magnetic moment observed for the Sr-based compound, induced by the increased proportion of low spin Co$^{3+}$, would augment this uncompensation and result in its largest ZEB effect. For La$_{1.5}$Ca$_{0.5}$CoMnO$_{6}$ (LCCMO) this AFM coupling is nearly compensated, resulting in a small $H_{EB}$.

\section{Experimental details}

Polycrystalline samples of LCMO, LBCMO, LCCMO and LSCMO  were synthesized by conventional solid state reaction, as described in the Supplementary Material (SM) \cite{SM}. The SXRD data were recorded at the XPD beamline of the Brazilian Synchrotron Light Laboratory (LNLS) using the Bragg-Brentano geometry. The anomalous scattering measurements were performed at room temperature using wavelength $\lambda$ = 1.9074 $\textrm{\AA}$, and the XRD patterns were obtained by one-dimensional Mythen-1K detector (Dectris). In order to further investigate the Co/Mn cationic order in LCMO sample, we also performed room temperature SXRD with $\lambda$ = 0.6525 $\textrm{\AA}$ at the x-ray diffraction and spectroscopy (XDS) beamline of LNLS \cite{XDS}. The low-$T$ measurements were performed at $\lambda$ = 1.3776 $\textrm{\AA}$ using a DE-202 cryostat (ARS Cryo) and the XRD patterns were obtained with a HOPG(002) analyser. The Rietveld refinements were performed using the program GSAS+EXPGUI \cite{GSAS}. 

The AC and DC magnetic measurements were carried out using a Quantum Design PPMS-VSM magnetometer. In order to prevent the presence of trapped current on the magnet and ensure a reliable ZFC process, from one measurement to another the samples were warmed up to the paramagnetic state and the coil was demagnetized in the oscillating mode.

Unpolarized Raman scattering measurements were performed at several $T$ on a Jobin Yvon T64000 triple 1800 mm$^{-1}$ grating spectrometer equipped with a liquid N$_{2}$-cooled multichannel CCD detector. The excitation was achieved with a 488 nm Ar$^{+}$ laser line in a quasi-backscattering configuration. Muon spin rotation and relaxation ($\mu$SR) experiments were performed at the Swiss Muon Source of Paul Scherrer Institut, Switzerland, using the nearly 100$\%$ spin-polarized positive muon beam at the GPS instrument. The measurements were performed for the LBCMO, LCCMO and LSCMO powder samples in zero field (ZF) and weak transverse field (wTF, field applied perpendicular to the initial muon spin direction) modes. The data were acquired at several $T$ between 5 and 300 K.  wTF experiments were performed under a field of  $H$ = 50 Oe.

XAS measurements at Co- and Mn-$K$ edges were performed at room temperature in the dispersive X-ray absorption (DXAS) beamline at LNLS \cite{DXAS}, and the Co- and Mn-$L_{3,2}$ XAS and XMCD spectra were recorded at room- and low-$T$ at the Planar Grating Monochromator (PGM) beamline of LNLS \cite{PGM}, in total electron yield (TEY) mode. The XAS and XMCD were investigated with the samples in the form of powder, but prior to the powdering of the bulk the pellet's surfaces were scraped in order to avoid the influence of its redox in the results. The circular polarization rate for the XMCD was of approximately 80\%.

\section{Results and discussions}

\subsection{Synchrotron X-ray diffraction}

The LCMO compound has been the subject of great academic interest since 1955, when J. B. Goodenough predicted FM insulating behavior for it \cite{Goodenough}. More recently, the interest was renewed with the discovery of room temperature magnetodielectric behavior \cite{Singh}, which is generally attributed to charge order of Co$^{2+}$ and Mn$^{4+}$ \cite{Chen,Blasco}. However, the presence of at least a small fraction of Co$^{3+}$ and Mn$^{3+}$ is frequently observed in bulk LCMO samples \cite{Dass,Burnus}, and there is an open debate on whether Co and Mn ions are ordered along the lattice or not \cite{Chen,Singh,Blasco,Goodenough,Burnus,Joy,Fournier}. The controversy takes place because the Co and Mn scattering factors are very alike, and consequently the conventional XRD results can be interpreted in terms of the monoclinic $P2_1/n$ space group, where Co and Mn order alternate along the lattice, as well as in terms of the orthorhombic $Pnma$ space group, corresponding to a disordered population of B-site. 

To address this issue we performed anomalous scattering measurements on LCMO using wavelength energy $E$ = 6500 eV. This energy was chosen to maximize the difference between the Co and Mn scattering factors. The small differences between the $P2_1/n$ and $Pnma$ diffraction patterns correspond to very weak reflections that are expected for the monoclinic symmetry, but not for the orthorhombic one. Long-lasting experiments were carried around the 2$\theta$ regions where these peaks are expected to appear, and there was no observation of such reflections. To further investigate that, we also performed a room temperature SXRD with $E$ = 19 keV and again no trace of these Bragg peaks associated to the $P2_1/n$ space group was observed (see SM \cite{SM}), giving thus strong evidence that our LCMO sample grows in $Pnma$ space group, corresponding nominally to a La(Co$_{0.5}$Mn$_{0.5}$)O$_{3}$ perovskite. Accordingly, all subsequent analysis of this sample considers it to present orthorhombic symmetry.

SXRD measurements with $E$ = 6500 eV were also performed for the Ba-, Ca- and Sr-doped compounds. For these samples the presence  of two crystallographic phases is clear, as already proposed for resemblant compounds \cite{Vashook,Taraphder,Xu,Berger}. For LCCMO, as observed for LCMO, it was not found the weak reflections expected for $P2_1/n$ space group. Therefore, the XRD pattern could be successfully refined with two $Pnma$ phases (94\% and 6\%) with almost equal cell volumes.  The fact that both LCMO and LCCMO belong to the same space group is expected, since Ca$^{2+}$ and La$^{3+}$ ionic radii are very close at XII coordination (1.34 and 1.36 \AA, respectively \cite{Shannon}). 

For LSCMO the SXRD data could be successfully refined with mixed rhombohedral $R\bar{3}c$ (91\%) and orthorhombic $Pnma$ (9\%) phases, while for LBCMO the Ba-doping brought the whole system to the rhombohedral phase, the crystalline structure being successfully calculated with two  $R\bar{3}c$ phases (96\% and 4\%). The phase segregation may be related to the formation of different domains rich in La-Ba/Ca/Sr and/or Co-Mn. Since the EB effect is generally attributed to the magnetic coupling at the interface of different magnetic phases, the presence of distinct crystallographic phases may be related to the ZEB observed in such materials.
 
\begin{figure}
\begin{center}
\includegraphics[width=0.47 \textwidth]{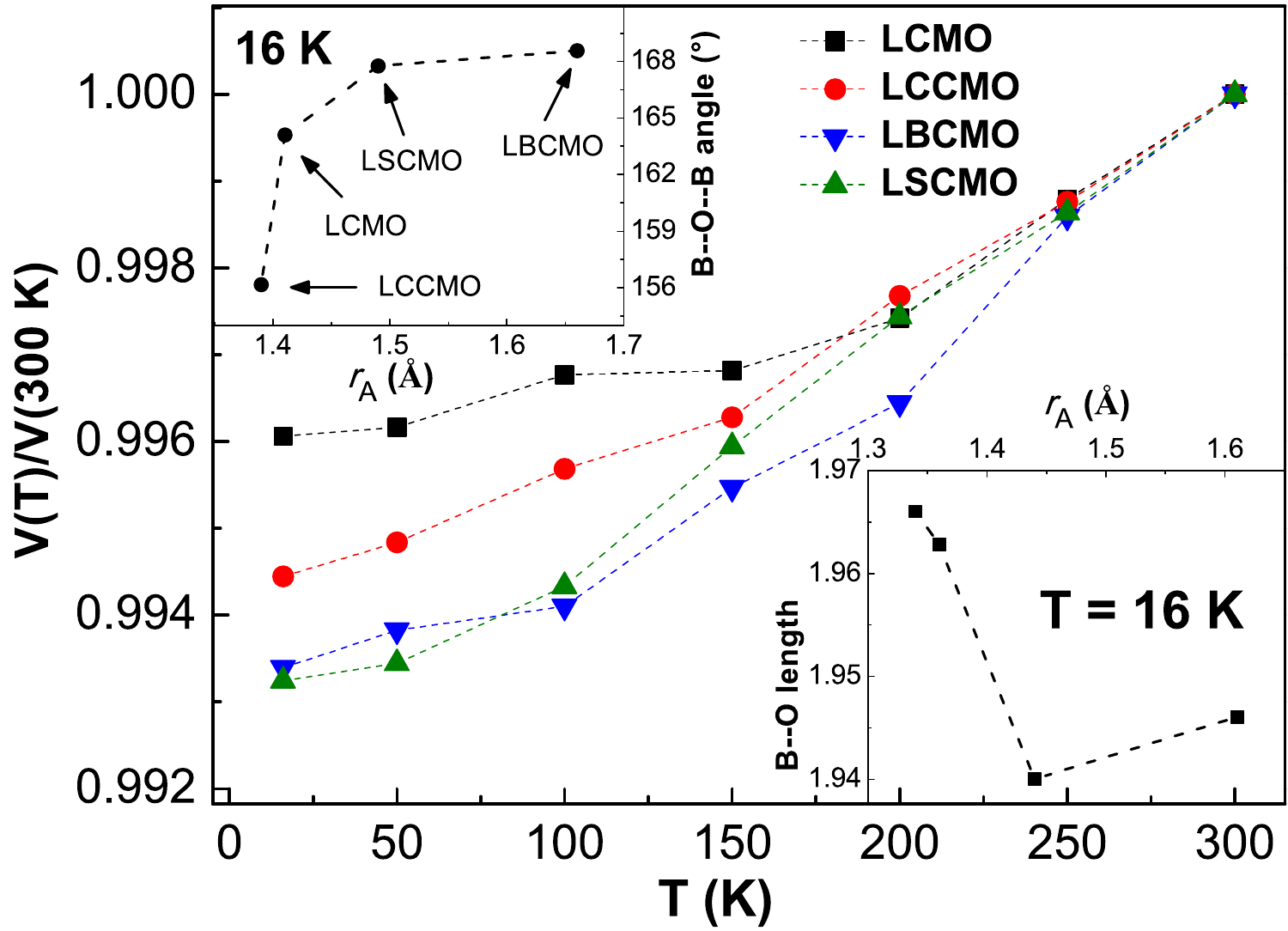}
\end{center}
\caption{Normalized average unit cell volume (V) as a function of $T$ for La$_{2-x}$A$_{x}$CoMnO$_{6}$ samples. The dashed lines are guides for the eye. The upper and bottom insets show respectively the average B-O bond length and B-O-B' bond angle as a function of the average A-site ionic radius, $r_A$, obtained from 16 K SXRD results.}
\label{Fig_SXRD}
\end{figure}

Since the ZEB effect is only observed at temperatures far below the magnetic transitions, we investigated the structural evolution of each sample with $T$ ranging from 300 down to 16 K, using wavelength $E$ = 9000 eV. Fig. \ref{Fig_SXRD} shows the evolution of the normalized average unit cell volume as a function of $T$. Although it was not observed structural transition for any compound, there are changes in the slope of the curves at temperatures close to the material's magnetic transitions, as will be discussed next. We can also note that the curves of LCMO and LCCMO are roughly similar, as well as those of LSCMO and LBCMO. These results are not surprising, since Ca$^{2+}$ and La$^{3+}$ ionic radii are close, whereas those for Ba$^{2+}$ and Sr$^{2+}$ are larger \cite{Shannon}.

The insets of Fig. \ref{Fig_SXRD} show the mean (Co/Mn)--O and Co--O--Mn bond lengths and angles obtained from the 16 K SXRD measurements. As expected, the Co--O--Mn bond angle increases as the A-site average ionic radius ($r_A$) increases \cite{Vasala,Woodward}. Interestingly, the mean B--O bond length presents the smaller value for LSCMO. Both B--O--B' angle and B--O length are strongly correlated to the magnetic coupling of transition metal (TM) ions in DP compounds \cite{Vasala,Serrate}. The closer to 180$^{\circ}$ is the B--O--B' angle, the more symmetric is the crystal structure, and the exchange interactions are stronger. On the other hand, in general the superexchange interaction gets stronger as the TM ions get closer. 

These results are certainly related to the magnetic properties observed for La$_{2-x}$A$_{x}$CoMnO$_{6}$. Despite the fact the B--O--B' angle is not the largest for LSCMO, Sr ionic radius is large enough to ensure a nearly symmetric structure for this sample, and the smaller B--O length ensures a strong coupling between Co and Mn, leading to the largest ZEB observed. This and other magnetic results will be discussed below. Although careful must be taken in interpreting these data since the oxygen scattering factor is small, the trends here observed are expected for DP compounds, and explain very well the magnetic properties of the system, being thus very plausible.

\subsection{AC and DC magnetization}

\begin{figure}
\begin{center}
\includegraphics[width=0.5 \textwidth]{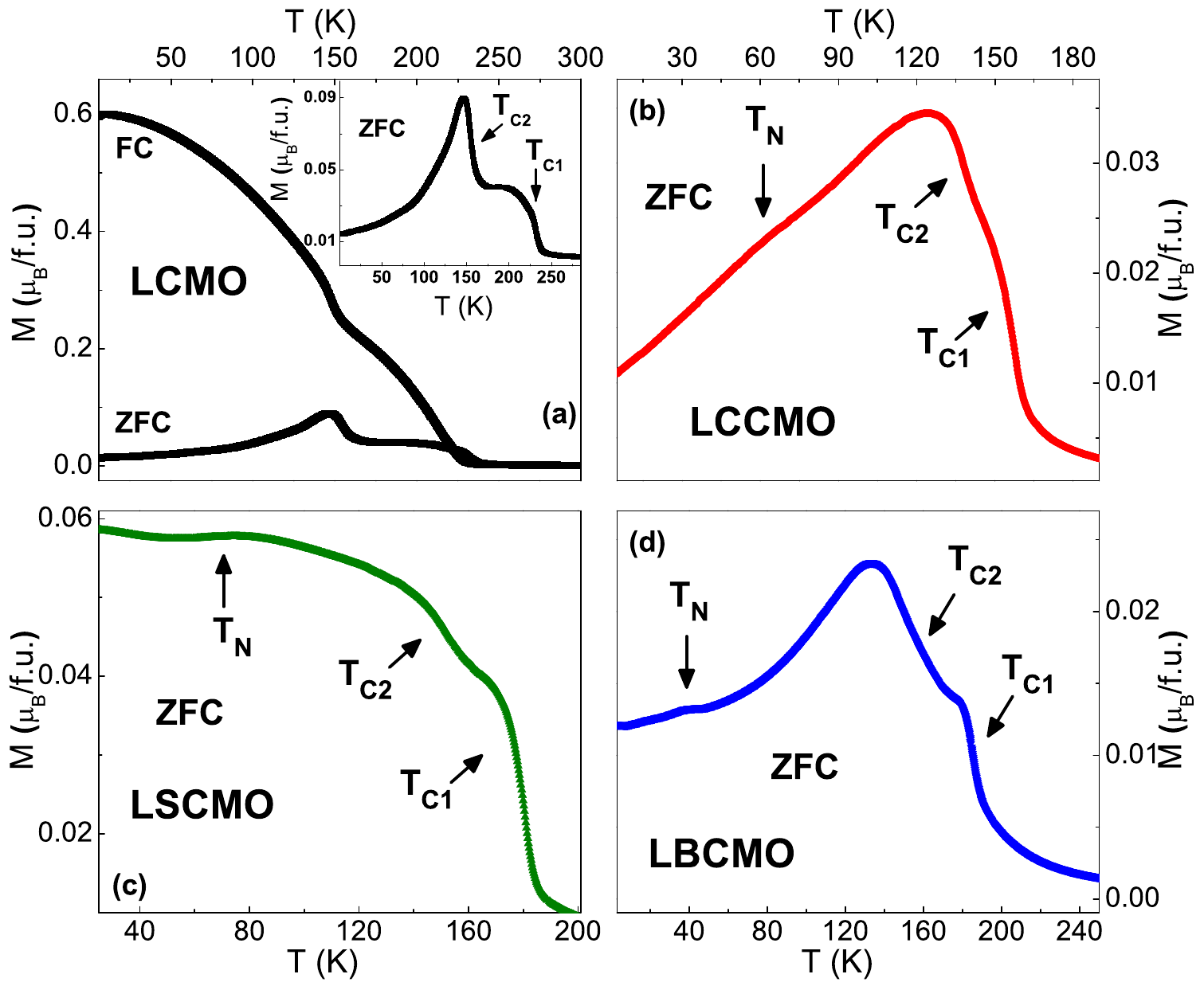}
\end{center}
\caption{(a) ZFC and FC $M(T)$ curves for LCMO, measured at $H$ = 100 Oe. The inset shows a magnified view of the ZFC curve, evidencing $T_{C1}$ and $T_{C2}$. (b), (c) and (d) show the ZFC $M(T)$ curves for LCCMO, LSCMO and LBCMO respectively.}
\label{Fig_MxT}
\end{figure}

Fig. \ref{Fig_MxT}(a) shows the $M(T)$ curves for LCMO, measured with $H$ = 100 Oe at 2 K$\leq$$T$$\leq$400 K. It can be clearly seen the presence of two anomalies, at $\sim$230  and $\sim$155 K. Early reports for this compound proposed Co$^{2+}$ and Mn$^{4+}$ configuration \cite{Goodenough}, but recently it is stablished that the Co and Mn valences are very sensitive to the conditions in which the sample is prepared \cite{Dass,Burnus}, the presence of Co$^{3+}$ and Mn$^{3+}$ being commonly observed in LCMO. The mixed valences Co$^{2+}$/Co$^{3+}$ and Mn$^{3+}$/Mn$^{4+}$ are confirmed by our XAS measurements (Section E) which, together with Goodenough-Kanamori-Anderson (GKA) rules \cite{GKA}, can plausibly explain the two anomalies observed in Fig. \ref{Fig_MxT}(a). According to GKA rules, Co$^{2+}$--O--Mn$^{4+}$ and Co$^{3+}$--O--Mn$^{3+}$ interactions are expected to be FM, the former coupling would correspond to $T_{C1}$=230 K transition while the latter would correspond to $T_{C2}$=155 K \cite{Chen,Burnus}. 

The $M(T)$ curves of Ba-/Ca-/Sr-doped samples are displayed in Figs. \ref{Fig_MxT}(b)-(d). It can be seen that the two peaks related to the FM transitions are present, although $T_{C1}$ is reduced. The replacement of 25\% of La$^{3+}$ by Ca$^{2+}$/Sr$^{2+}$/Ba$^{2+}$ changes the electronic configuration of the TM ions. For the doped compounds an increase of the average Co and/or Mn valence in relation to LCMO is expected, signifying a decrease in the concentration of the Co$^{2+}$--O--Mn$^{4+}$ coupling, and this may be directly related to the decrease of their ordering $T'$s.

From the fit to the magnetic susceptibility curves in the paramagnetic region with the Curie-Weiss law we obtained the Curie-Weiss temperature, $\theta_{CW}$, and the effective magnetic moment, $\mu_{eff}$ (see table \ref{Tmag}). For all investigated samples $\theta_{CW}$ is positive and fairly close to $T_{C1}$, giving further evidence of the prominent FM coupling. For LCMO $\mu_{eff}$=6.7 $\mu_{B}$/f.u., in agreement with previous reports \cite{Blasco,Dass,Joy}. For Ba- and Sr-doped samples it decreases to 6.6 and 6.2 $\mu_{B}$/f.u., respectively, which could be related to the increased proportion of Mn$^{4+}$ and/or Co$^{3+}$ in the low spin (LS) configuration, as will be discussed in section E. Interestingly, for the Ca-based sample there is no significant decrease of $\mu_{eff}$ in relation to the parent compound, which could indicate that this system is favorable for the emergence of high spin (HS) or intermediate spin (IS) Co$^{3+}$. Nevertheless, these results are certainly related to the different ZEB effects observed in these samples. 

\begin{table}
\renewcommand{\arraystretch}{1.2}
\caption{Main results obtained from the AC and DC $M(T)$ curves, and from the ZFC $M(H)$ measurements.}
\label{Tmag}
\resizebox{\columnwidth}{!}{
\begin{tabular}{ccccc}
\hline \hline
Sample & LCMO & LCCMO & LSCMO & LBCMO \\
\hline
$T_{C1}$ (K) & 230 & 158 & 180 & 186 \\

$T_{C2}$ (K) & 157 & 141 & 157 & 155 \\

$T_{N}$ (K) & - & 62 & 74 & 45 \\

$\theta_{CW}$ (K) & 213 & 185 & 220 & 187 \\

$\mu_{eff}$ ($\mu_B$/f.u.) & 6.7 & 6.7 & 6.2 & 6.6 \\

$T_g$ (K) & - & 46.7 & 72.9 & 71.7 \\

$\tau_0$ (s) & - & 3.9$\times$10$^{-7}$ & 3.9$\times$10$^{-7}$ & 2.8$\times$10$^{-8}$ \\

\textit{z}$\nu$ & - & 8.9 & 5.8 & 6.5 \\

$\delta T_{f}$ & - & 0.07 & 0.09 & 0.08 \\

$H_C$ (Oe) & 7694 & 8437 & 7147 & 5178 \\

$H_{EB}$ (Oe) & - & 253 & 3128 & 1605 \\
\hline \hline
\end{tabular}}
\end{table}

It is important to note the appearance of subtle anomalies at the low-$T$ regions of the $M(T)$ curves of Ba-/Ca-/Sr-doped samples. This may be related to the emergence of non-negligible portions of other magnetic interactions like Co$^{3+}$--O--Mn$^{4+}$ and Co$^{2+}$--O--Co$^{3+}$, which are predicted by GKA rules to be AFM. The presence of competing magnetic phases, as well as disorder, are key ingredients to the emergence of SG-like behavior. In order to verify the presence of SG-like phase, which is believed to play an important role on the ZEB effect of DP systems \cite{ZEBmodel}, we performed AC magnetic susceptibility measurements ($\chi_{AC}$) at several frequencies ($f$) in the range 25-10000 Hz, using $H_{AC}$ = 5 Oe. For LCMO it was not found evidence of glassy magnetism\cite{SM}, as expected. This compound was exhaustively investigated in the last decades and in general SG-like behavior is not suggested.

For the Ba-/Ca-/Sr-doped samples one can clearly observe a monotonic increase of the freezing $T$ ($T_f$) with increasing $f$ (see SM \cite{SM}), suggesting SG-like behavior.  The $T_f$ as a function of $f$ curves of each sample could be well fitted by the power law equation of the dynamic scaling theory, commonly used to investigate SG-like systems \cite{Mydosh,Souletie}
\begin{equation}
\frac{\tau}{\tau_{0}}=\left[\dfrac{(T_{f} - T_{g})}{T_{g}}\right]^{-z\nu} \label{Eq1}
\end{equation}
where $\tau$ is the relaxation time corresponding to the measured frequency, $\tau_{0}$ is the characteristic relaxation time of spin flip, $T_{g}$ is the SG-like transition temperature ($T_f$ as $f$ tends to zero), $z$ is the dynamical critical exponent and $\nu$ is the critical exponent of the correlation length. The main results obtained from the fittings are displayed in table \ref{Tmag}, where it can be noted that the $\tau_{0}$ and $z\nu$ values are typically found for cluster spin glass (CG) systems \cite{Souletie,Malinowski,Murthy2}.

Usually, a formula computing the shift in $T_{f}$ between the two outermost frequencies is used to classify the material as SG, CG or superparamagnet (SP)
\begin{equation}
\delta T_{f}=\frac{\triangle T_{f}}{T_{f} \triangle log f}. \label{Eq2}
\end{equation}
The $\delta T_{f}$ values obtained for the doped samples (see Table \ref{Tmag}) lie in the range $0.01\lesssim\delta T_{f}\lesssim0.1$ usually found for CG systems. For canonical SG, the usual value is $\delta T_{f}\lesssim0.01$, while for SP $\delta T_{f}\gtrsim0.1$ \cite{Souletie,Malinowski,Murthy2,Anand2}. These results, together with the conventional magnetic transitions observed at higher $T$, confirms the RSG state on doped La$_{1.5}$A$_{0.5}$CoMnO$_{6}$ compounds that present ZEB effect, while for the parent LCMO compound there is no RSG nor ZEB behavior.

\begin{figure*}
\begin{center}
\includegraphics[width= \textwidth]{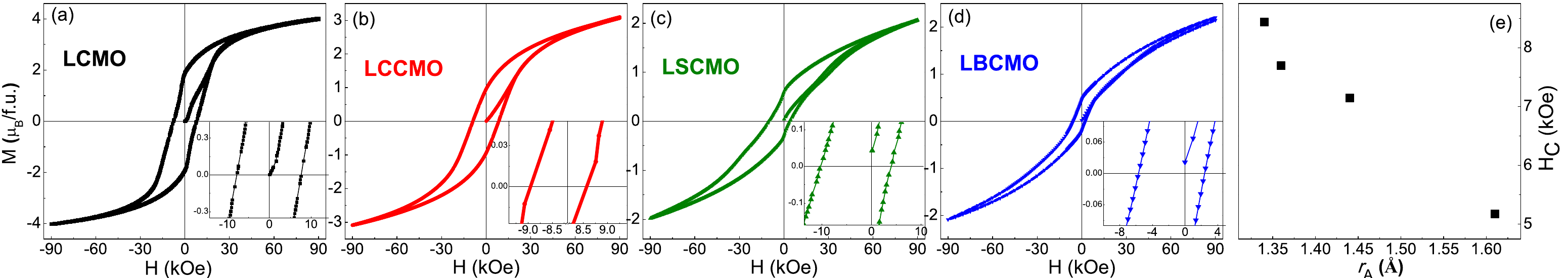}
\end{center}
\caption{ZFC $M(H)$ curves of (a) LCMO, (b) LCCMO, (c) LSCMO and (d) LBCMO, measured at $T$ = 5 K and $H_{max}$ = 90 kOe. The insets show magnified views of the curves close to the $M$ = 0 region, evidencing the ZEB effect in the doped samples. (d) $H_C$ as a function of $r_A$.}
\label{Fig_MxH}
\end{figure*}

In order to verify the ZEB effect on the La$_{2-x}$A$_{x}$CoMnO$_{6}$ system, we performed $M(H)$ measurements after ZFC each sample. Fig. \ref{Fig_MxH} shows the curves obtained at $T$ = 5 K, using a maximum applied field $H_{max}$ = 90 kOe. For all samples we observe closed loops, fairly symmetric with respect to the $M$ axis. For LCMO there is no EB, \textit{i.e.}, the $M(H)$ curve is also symmetric with respect to $H$ axis. As commented, this compound was extensively investigated and, to the best of our knowledge, EB effect was never reported for LCMO. It is interesting to note the lack of a complete saturation, and the $M$ $\sim$4.1 $\mu_{B}$/f.u. observed at $H_{max}$ = 90 kOe is fairly below the value expected for a FM system containing Co$^{2+}$/Co$^{3+}$ and Mn$^{3+}$/Mn$^{4+}$. These results were previously explained in terms of the appearance of Mn--O--Mn and Co--O--Co AFM bonds, the presence of LS Co$^{3+}$ and the formation of antiphase boundaries, that leads to the antiparallel coupling of neighboring domains \cite{Blasco,Dass,Fournier}.

The $M(H)$ curves of the doped compounds are shifted toward left along the $H$ axis, characterizing the ZEB effect. The EB and the coercive fields are respectively defined as $H_{EB}=|H^{+}+H^{-}|/2$ and $H_{C}=(H^{+}-H^{-})/2$, where $H^{+}$ and $H^{-}$ are the positive and negative coercive fields, respectively. The Sr-based sample presents the largest ZEB effect reported so far, $H_{EB}$ $\sim$ 3100 Oe at 5 K. The Ba-based compound also exhibits a pronounced shift, $H_{EB}$ $\sim$ 1600 Oe, while the Ca-based sample, on the other hand, presents a very small shift, $H_{EB}$ $\sim$ 250 Oe. These results are certainly related to the structural and electronic configuration of each sample. As discussed above, LSCMO presents a large B--O--B' angle and the smallest B--O length. It is also interesting to note the decrease of the ``saturation" magnetization ($M_{sat}$) of the doped samples in relation to the parent compound. This can be understood in terms of the increased proportion of LS Co$^{3+}$ induced by Ca$^{2+}$/Ba$^{2+}$/Sr$^{2+}$ to La$^{3+}$ substitution, as will be discussed in Section E. This decrease is more pronounced for Sr- and Ba-based samples than for the Ca-based one, being certainly related to the small $H_{EB}$ observed for the last compound.

The $M(H)$ curves also show an interesting decrease of $H_{C}$ with the increase of the average A-site ionic radius ($r_A$), Fig. \ref{Fig_MxH}(e). Previous studies of chemical pressure on DP compounds report the increase of $H_{C}$ with the decrease of $r_A$, which was attributed to the enhancement of the orbital magnetic moment of the TM ions caused by the increased lattice distortion \cite{Sikora,Haskel}. These results keep resemblance with those obtained for our system. As will be shown in section E, the Co relative orbital-to-spin moment ratio, $m_L$/$m_S$, is greatly reduced in LSCMO in comparison to LCCMO. Although the $m_L$/$m_S$ ratio is even larger for LCMO than for LCCMO, the average $r_A$ is smaller for the later. Recent investigations of applied external pressure on DP compounds have shown that, even when the physical pressure does not significantly alter the orbital moment, it drastically changes the coercivity, being this related to pressure-induced changes on the crystal field \cite{Haskel}. In the case of La$_{2-x}$A$_{x}$CoMnO$_{6}$, the combined effect of crystal field and spin-orbit interaction seems to play a role on the magnetic anisotropy.

\subsection{Muon rotation and relaxation}

\begin{figure}
\begin{center}
\includegraphics[width=0.42 \textwidth]{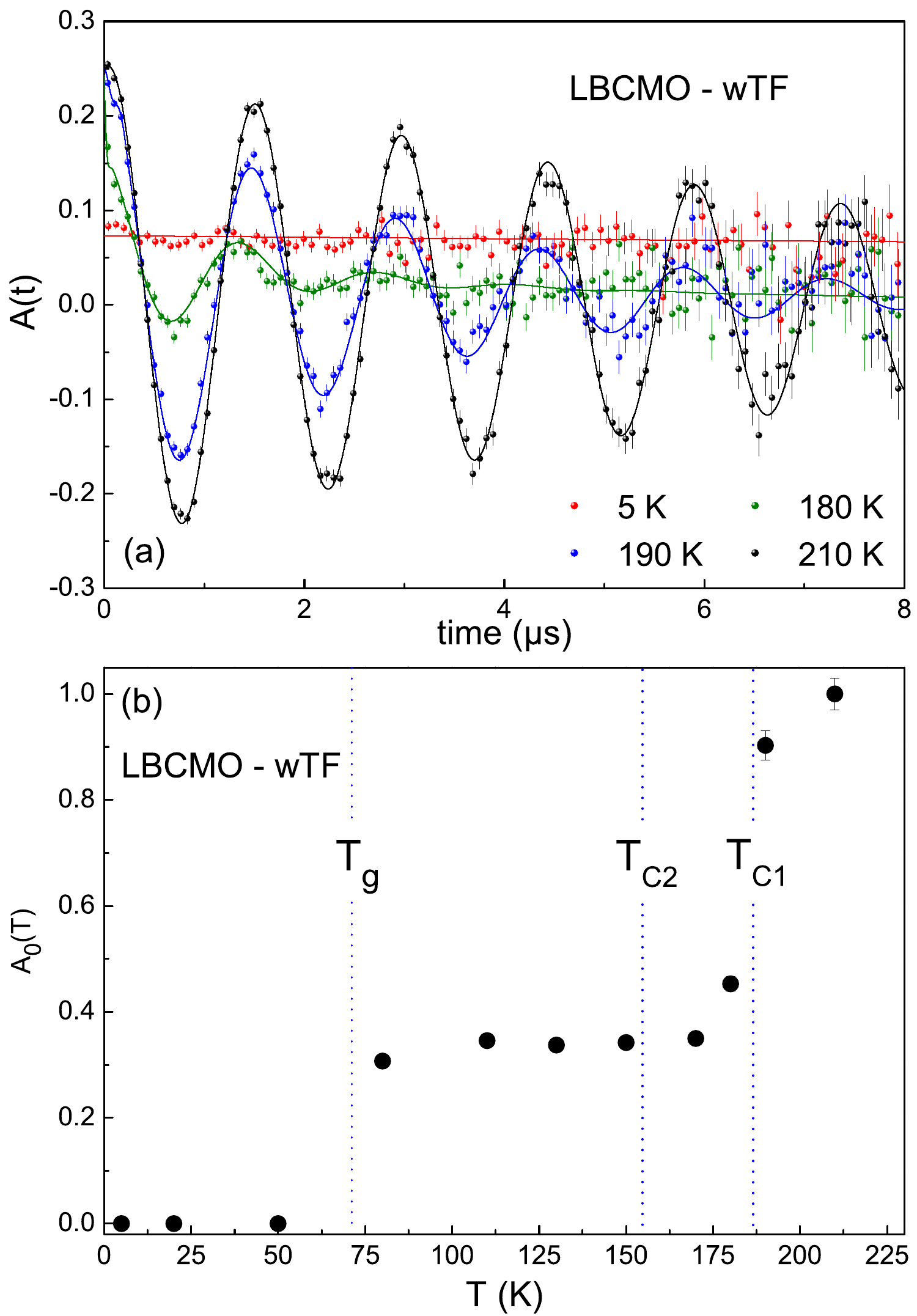}
\end{center}
\caption{(a) $\mu$SR rotation patterns of LBMCO in a weak applied transverse field (wTF) of 50 Oe at various temperatures; (b) variation of weakly damped paramagnetic fraction with temperature derived from wTF spectra. Indicated marks for $T_{C1}$, $T_{C2}$ and $T_g$ are taken from magnetization and susceptibility data.}
\label{Fig_muSR_TF}
\end{figure}

As discussed in the previous section, the presence of two FM couplings in CoMn-based DPs is well known, being attributed to Co$^{2+}$--O--Mn$^{4+}$ and Co$^{3+}$--O--Mn$^{3+}$ interactions \cite{GKA,Chen,Burnus}. However, the ZFC $M(T)$ experiments revealed the presence of a third anomaly at lower-$T$ for the Ba-, Ca- and Sr-doped compounds. For further clarification of the magnetic properties we performed $\mu$SR experiments on these samples. Representative spectra taken in wTF mode at several temperatures for LBCMO are shown in Fig. \ref{Fig_muSR_TF}(a).

In wTF experiments, the onset of magnetic ordering causes an apparent loss of asymmetry of the initially 100$\%$ polarized muon spins due to randomly adding strong local magnetic fields to the applied magnetic field causing a severe damping of the signal. Comparing the spectra taken at 210 K and 180 K, one can clearly observe changes in the initial asymmetry of a weakly damped rotation signal, confirming that the first magnetic transition $T_{C1}$ is within this $T$ range, as found in the $M(T)$ curves. Already at 190 K one can detect a small contribution by a spontaneous rotation seen at very early times indicating a partial onset of magnetic order.  The dominant signal represents a rotation of muon spins with a frequency $\omega_{\mu}$. This signal can be fitted to 
\begin{equation}
A(t) = A(t=0)\cdot A_{0}(t)\cdot G_{par}(t)\cdot cos(\omega_{\mu}t), \label{Eq3}
\end{equation}
where $\omega_{\mu}$ = $\gamma_{\mu}B_{\mu}$, with $\gamma_{\mu}$ = 2$\pi\cdot$135.54 $\mu$s$^{-1}T^{-1}$ being the muon gyromagnetic ratio and $B_{\mu}$ the local field at the muon site. $G_{par}(t)$ = $e^{-\lambda t}$ is the depolarization function with a damping factor $\lambda$ due to field inhomogeneity and relaxation. Notably, $B_{\mu}$ is slightly larger than the applied field and increasing upon lowering $T$. This is due to a Knight shift adding to the applied field related to a $T$ dependent susceptibility. The rotating signal comes from a paramagnetic (PM) fraction still existing below the FM transitions. In Fig. \ref{Fig_muSR_TF}(b) we have plotted the variation of this PM fraction with $T$. Between $T_{C2}$ down to $T_g$ there is a nearly constant PM fraction of about 35\%, at lower $T$ the PM fraction vanishes.

\begin{figure}
\begin{center}
\includegraphics[width=0.42 \textwidth]{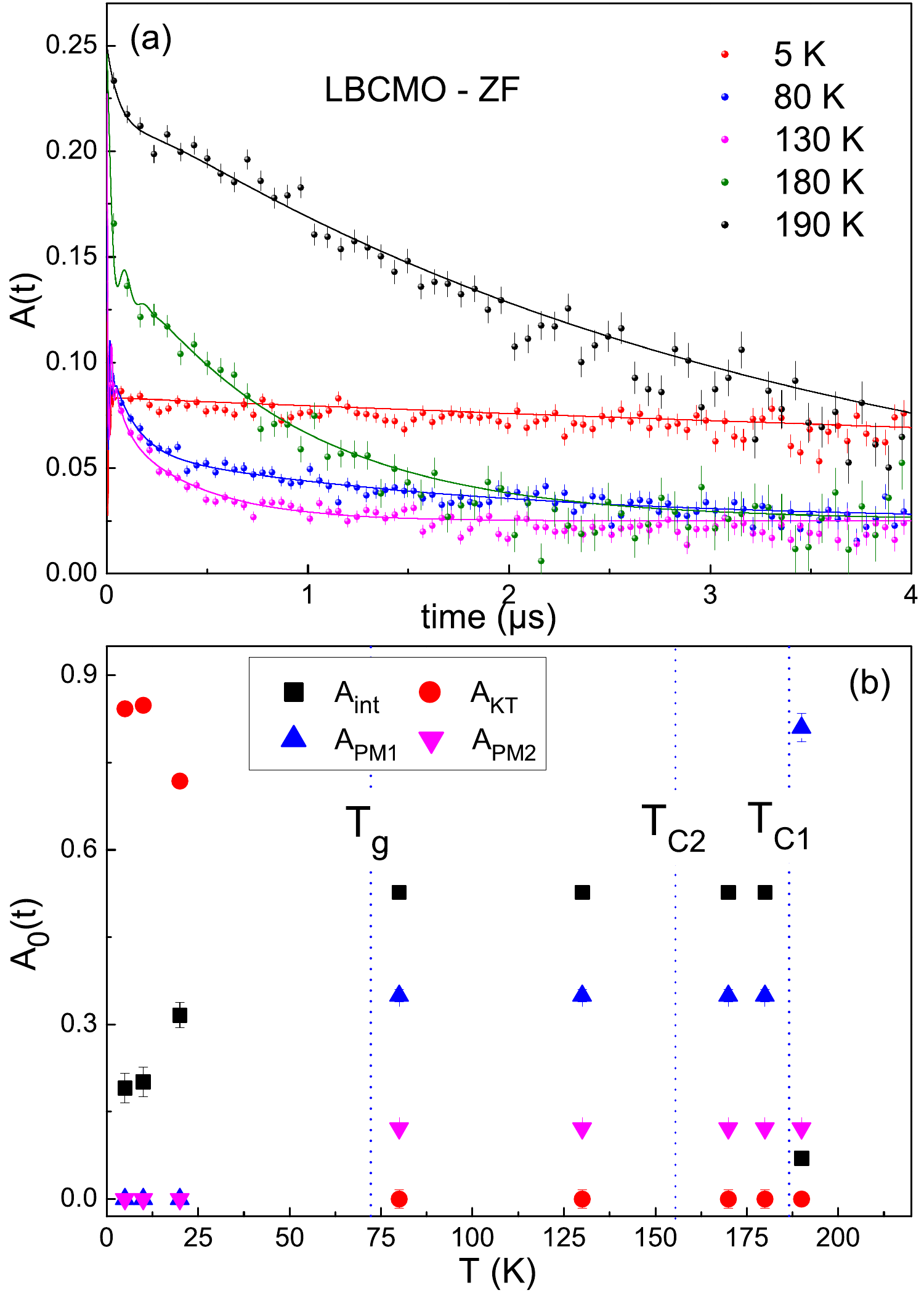}
\end{center}
\caption{(a) ZF $\mu$SR of LBCMO for various $T$; (b) $T$ dependence of partial asymmetries. Indicated marks for $T_{C1}$, $T_{C2}$ and $T_g$ are taken from magnetization and susceptibility data.}
\label{Fig_muSR_ZF}
\end{figure}

The wTF data are supported by ZF $\mu$SR spectra, Fig. \ref{Fig_muSR_ZF}(a) (see also SM \cite{SM}). Fits to these spectra suggest the presence of several superimposed signals with different temperature dependent partial asymmetries summing up to a total observed asymmetry $A(t)$:
\begin{equation}
\begin{split}
A(t) & = A(t=0)\{A_{PM1}\cdot G_{PM1}(t)+A_{PM2}\cdot G_{PM2} \\
& + A_{int}\cdot G_{int}(t)+A_{KT}\cdot G_{KT}(t)\}. \label{Eq4} \\
\end{split}
\end{equation}
As will be discussed below in more detail, $A_{PM1}$, $A_{PM2}$, $A_{int}$, and $A_{KT}$ are the partial asymmetry contributions from a slowly exponentially damped paramagnetic, a stretched exponentially damped paramagnetic, an internal field and a Gaussian Kubo-Toyabe signal, respectively. $G(t)$ are the corresponding depolarization factors that are explicitly given in SM \cite{SM}.

In Fig. \ref{Fig_muSR_ZF}(b) we have plotted the $T$ dependence of the different partial contributions to the spectra. The higher sensitivity of ZF spectra to different damping behavior of subspectra allows a more detailed analysis. The onset of spontaneous rotations below $T_{C1}$ and the variation of the local field at the muon sites can be clearly traced especially for the early times of spectra (see SM \cite{SM}). Down to about $T_g$ we see the spontaneously rotating signal with a frequency $\omega_{int}$ = $\gamma_{\mu}B_{int}$ having a spectral contribution $A_{int}$ with about 50\% of total asymmetry (third term of Eq. \ref{Eq4}); this fraction is long-range ordered. The $T$ dependent magnetic field at the muon site $B_{int}$ is plotted in SM \cite{SM}.  

In correspondence with the wTF spectra, there is a slowly exponentially damped paramagnetic signal (PM1) having a spectral fraction of about 35\%  (first term in Eq. \ref{Eq4}) . In addition there is, however, a contribution of about 10\% that is best fitted with a stretched exponential depolarization function $e^{-(\lambda_{2}t)^{\beta}}$ (second term in Eq. \ref{Eq4}) typical for a wide distribution of fluctuation frequencies as expected for a still dynamically fluctuating, but gradually freezing spin glass. Values of the exponent $\beta$ vary from 1 to about 0.4 between $T_{C2}$ and $T_g$. The damping parameter $\lambda_2$ is much larger than for PM1, indicating longer correlation times for PM2 and, for this reason, it was not possible to resolve this contribution from wTF. 

Below $T_g$ the relative contributions from subspectra change. The PM1 fraction vanishes and  the spontaneously rotating signal decreases. About 80\% of the spectrum can be described by a so-called static Gaussian Kubo-Toyabe function (last term in Eq. \ref{Eq4})
\begin{equation}
G_{KT}(t) = \frac{1}{3} + \frac{2}{3}[1-(\sigma t)^2]e^{-(\sigma t)^{2}/2}, \label{Eq5}
\end{equation}
that is typical for a randomly frozen spin system producing a field distribution with a Gaussian width $\sigma$ at the muon site.

For $T$ $\geq$ 80 K the time dependent asymmetries in Fig. \ref{Fig_muSR_ZF}(a) reveal a continuous decrease up to long times due to relaxational damping and a levelling off with about 10\% of initial asymmetry. This corresponds to the muon spins in an internal field oriented parallel or antiparallel to the initial polarization of the muon beam. Therefore these muon spins do not undergo spontaneous rotation in the internal field, instead they give rise to a so-called 1/3 tail (see SM \cite{SM}). In our case this is only very weakly damped, \textit{i.e.} magnetic fluctuations are nearly absent at these muon sites. Note, that this 1/3 tail increases when going below $T_g$, since now we have also an additional 1/3 tail from the static Kubo-Toyabe contribution (first term in Eq. \ref{Eq5}).  

For the Ca- and Sr- doped samples we found very similar results (see SM \cite{SM}). Since the $M(T)$ curves did not reveal a low-$T$ anomaly for the parent compound LCMO, and as will be shown in section E, the Ba/Ca/Sr doping leads to the enhancement of Co$^{3+}$, the vanishing of the PM1 fraction at 50 K may be associated to the onset of a third magnetic transition related most probably to the Co$^{3+}$--O--Mn$^{4+}$ AFM coupling. Frustration is leading to a SG-like state, as clearly seen by ZF $\mu$SR, in agreement with magnetic susceptibility data.

\subsection{Raman spectroscopy}

\begin{figure}
\begin{center}
\includegraphics[width=0.5 \textwidth]{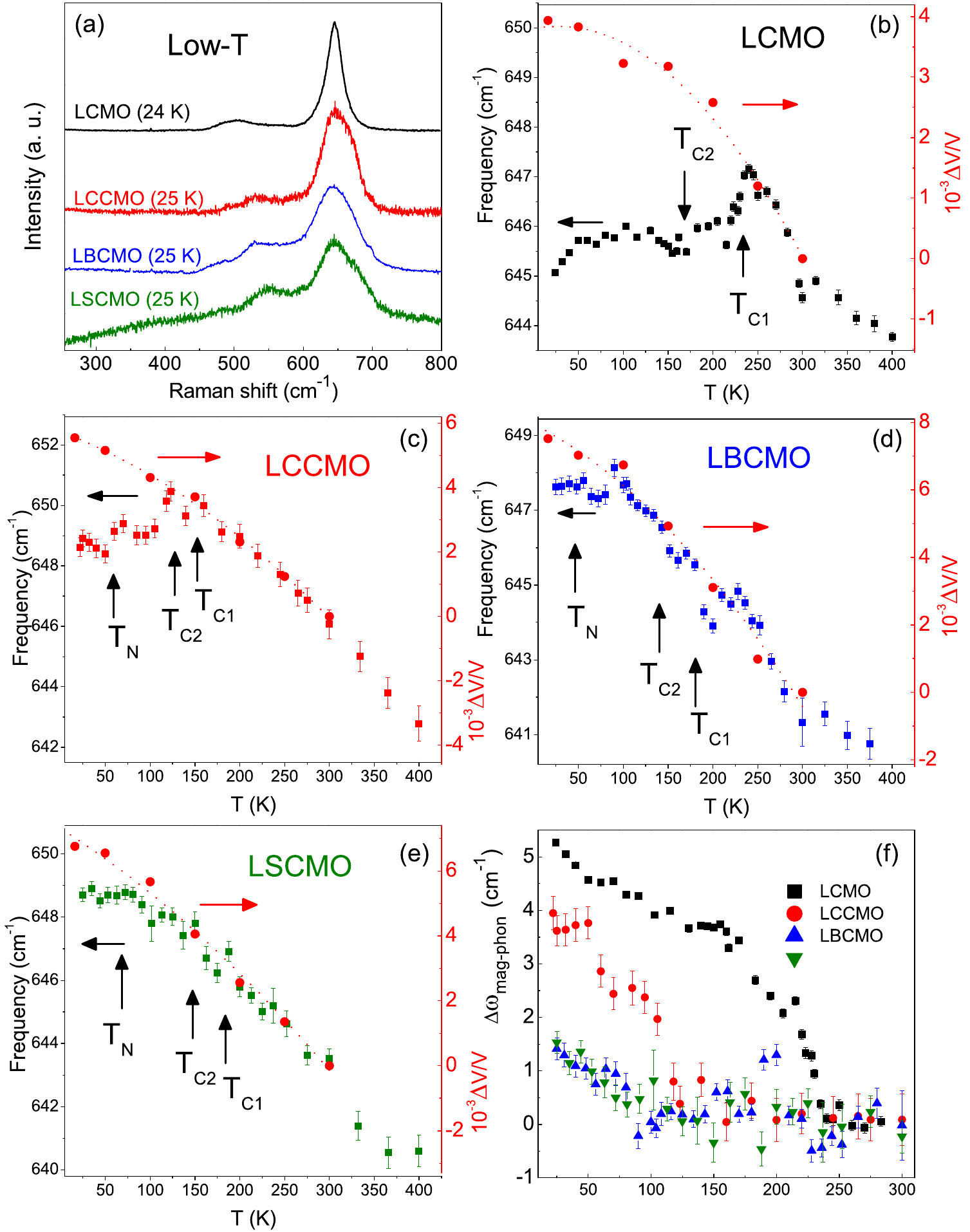}
\end{center}
\caption{(a) Low-$T$ Raman spectra of La$_{2-x}$A$_{x}$CoMnO$_{6}$ samples. The $T$-dependences of the relative shifts in the stretching mode are displayed for (b) LCMO, (c) LCCMO, (d) LBCMO and (e) LSCMO, where the solid circles show the relative unit cell volume variations extracted from SXRD data. The doted lines are guides to the eye. (f) shows the $\triangle\omega_{mag-phon}(T)$ for all investigated samples (see text).}
\label{Fig_Raman}
\end{figure}

Raman spectroscopy is a sensitive and powerful tool to investigate charge/orbital ordering and spin-lattice interactions in perovskite compounds. Unpolarized Raman spectra were taken at several $T$ in the range 24-400 K for the four investigated samples. As can be seen in Fig. \ref{Fig_Raman}(a), the observed spectra are consistent with previous works reported for LCMO and resemblant compounds, where a broad peak at $\sim650$ cm$^{-1}$ corresponding to the symmetric stretching mode and another at $\sim500$ cm$^{-1}$ associated with either anti-stretching or bending mode vibrations of (Co/Mn)O$_{6}$ octahedra can be noticed \cite{Fournier,Murthy2,Iliev,Murthy3,Silva}. Due to the possible overlapping of Raman-active modes below 600 cm$^{-1}$, we focused here on the $\sim650$ cm$^{-1}$ stretching mode. A detailed description of the anti-stretching/bending mode can be found in SM \cite{SM}.

To obtain the phonon parameters as a function of $T$, the observed peaks were fitted with Lorentzian lineshapes, which are good approximations for the classical Raman response from damped harmonic oscillators \cite{Cardona}. Figs. \ref{Fig_Raman}(b)-(e) show the $T$-dependence of the Raman shift for the stretching mode. The resulting curves show characteristic changes in the Raman shift that are closely related to each material's magnetic transitions. The figures also show a large softening of the phonon frequencies for LCMO and LCCMO below $T_C$, while for LBCMO and LSCMO this tendency is not so clear, although changes in the slope of the curves still seems to be present at $T$ $\sim$ $T_C$.

Both the magnetic couplings and the structural changes observed for these samples are expected to impact the phonon frequency. The changes in the phonon frequency due to thermal lattice expansion can be expressed by the following equation \cite{Gervais}
\begin{equation}
\delta \omega_{n}(T)= -\omega_{n}\int_{0}^{T}g_{n}(T)\alpha_{V}(T)dT, \label{Eq6}
\end{equation}
where $n$ refers to the $n$-th phonon mode, $\omega_{n}$ is its frequency, $g_{n}(T)$ is its Gr\"{u}neisen parameter and $\alpha_{V}(T)$ is the volumetric thermal expansion coefficient. For the $T$-range here investigated, $g_{n}$ can be assumed to be nearly $T$-independent, resulting in
\begin{equation}
\triangle \omega_{n}(T)= -\omega_{n}g_{n}\frac{\triangle V}{V}. \label{Eq7}
\end{equation}
Figs. \ref{Fig_Raman}(b)-(e) compares the evolution of -$\Delta$V/V obtained from the SXRD data with the phonon frequencies, confirming that the tendency of softening of the phonon frequency at low-$T$ is most likely related to the magnetic orderings. In previous reports for LCMO this softening was interpreted as a signature of spin-lattice coupling, which in turn was associated to the magnetodielectric effect observed for this compound \cite{Iliev,Murthy3}. In order to further investigate this possible spin-lattice coupling, we have plotted in Fig. \ref{Fig_Raman}(f) the changes in the phonon frequency of the stretching mode with the contribution of the structural variations discounted, \textit{i.e.}, we have used the following equation
\begin{equation}
\triangle\omega_{mag-phon}(T)= \triangle \omega_{n}-\triangle \omega(T), \label{Eq8}
\end{equation}
where $\triangle\omega(T)$ represents the $\triangle\omega$ obtained from the Raman spectra  and $\triangle \omega_{n}$ corresponds to the changes in the phonon frequency due to the structural changes, extracted from Eq. \ref{Eq6} (see also Ref. \citenum{Granado}). The resulting curves indicate a strong spin-phonon coupling for LCMO below $T_C$ and a smaller, but still significant, effect for LCCMO. For LBCMO and LSCMO the effect is greatly reduced, but non-negligible $\triangle\omega_{mag-phon}$ values can be noticed at low-$T$. A detailed investigation of the material's dielectric properties is necessary to verify the possible magnetodielectric effect in the Ba-,Ca- and Sr-doped compounds.

\begin{table}
\caption{FWHD obtained from the Lorentzian fits of Raman spectra carried at 300 K and 25 K for all samples, except for LCMO whose spectra were taken at 296 K and 24 K.}
\label{Traman}
\begin{tabular}{c|c|c|c|c}
\hline \hline
Sample & \multicolumn{4}{c}{FWHM (cm$^{-1}$)} \\
\hline
  & \multicolumn{2}{c|}{Room temperature} & \multicolumn{2}{c}{Low temperature}\\
\hline
  & stretching & anti-stretching & stretching & anti-stretching \\
\hline
LCMO & 40.3(2) & 80.1(26) & 26.7(1) & 79.5(20) \\
LCCMO & 63.8(9) & 113.6(87) & 52.8(6) & 88.6(68) \\
LBCMO & 97.5(20) & 70.9(77) & 70.2(8) & 75.1(30) \\
LSCMO & 81.6(13) & 126.0(73) & 74.1(9) & 100.3(30) \\
\hline \hline
\end{tabular}
\end{table}

Comparing the Raman spectra of Fig. \ref{Fig_Raman}(a), it is interesting to observe that the anti-stretching/bending peaks of the doped ZEB samples are shifted to higher frequencies in relation to that of non-ZEB LCMO. This monotonic dislocation of the peak position follows the same trend of increase of $H_{EB}$. It can also be noticed a pronounced broadening of the stretching mode of the doped samples in relation to the parent compound (see also Table \ref{Traman}), most likely related to the increase of disorder and/or the presence of biphasic crystal structure in the samples \cite{Fournier,Murthy2}. In the case of our Ba-, Ca- and Sr-doped samples there are both cationic disorder and the presence of two structural phases. Table \ref{Traman} shows that LSCMO presents the broader peaks at low-$T$. It is known that disorder and competing magnetic phases are key ingredients to the appearance of SG-like behavior, which in turn is necessary for the emergence of ZEB effect \cite{ZEBmodel}. The higher inhomogeneity observed for LSCMO in the SXRD and Raman data may be related to its larger ZEB.
 
\subsection{X-ray absorption spectroscopy}

XAS spectra at $L_{2,3}$-edges of TM ions are very sensitive to valence states. In the case of first row TM ions, different valences produce clearly differentiated final states in the 2$p^{6}$3$d^{n}$ to 2$p^{5}$3$d^{n+1}$ absorption process which translate into shifts in the energy position of the absorption peaks of the spectra. In order to determine the Co- and Mn-valence states in the La$_{2-x}$A$_{x}$CoMnO$_{6}$ samples, we performed XAS measurements at Co- and Mn-$L_{2,3}$ edges. We used CoO, LaCoO$_3$, LaMnO$_3$ and CaMnO$_3$ as reference samples for Co$^{2+}$, Co$^{3+}$, Mn$^{3+}$ and Mn$^{4+}$ configuration, respectively. 

\begin{figure}
\begin{center}
\includegraphics[width=0.5 \textwidth]{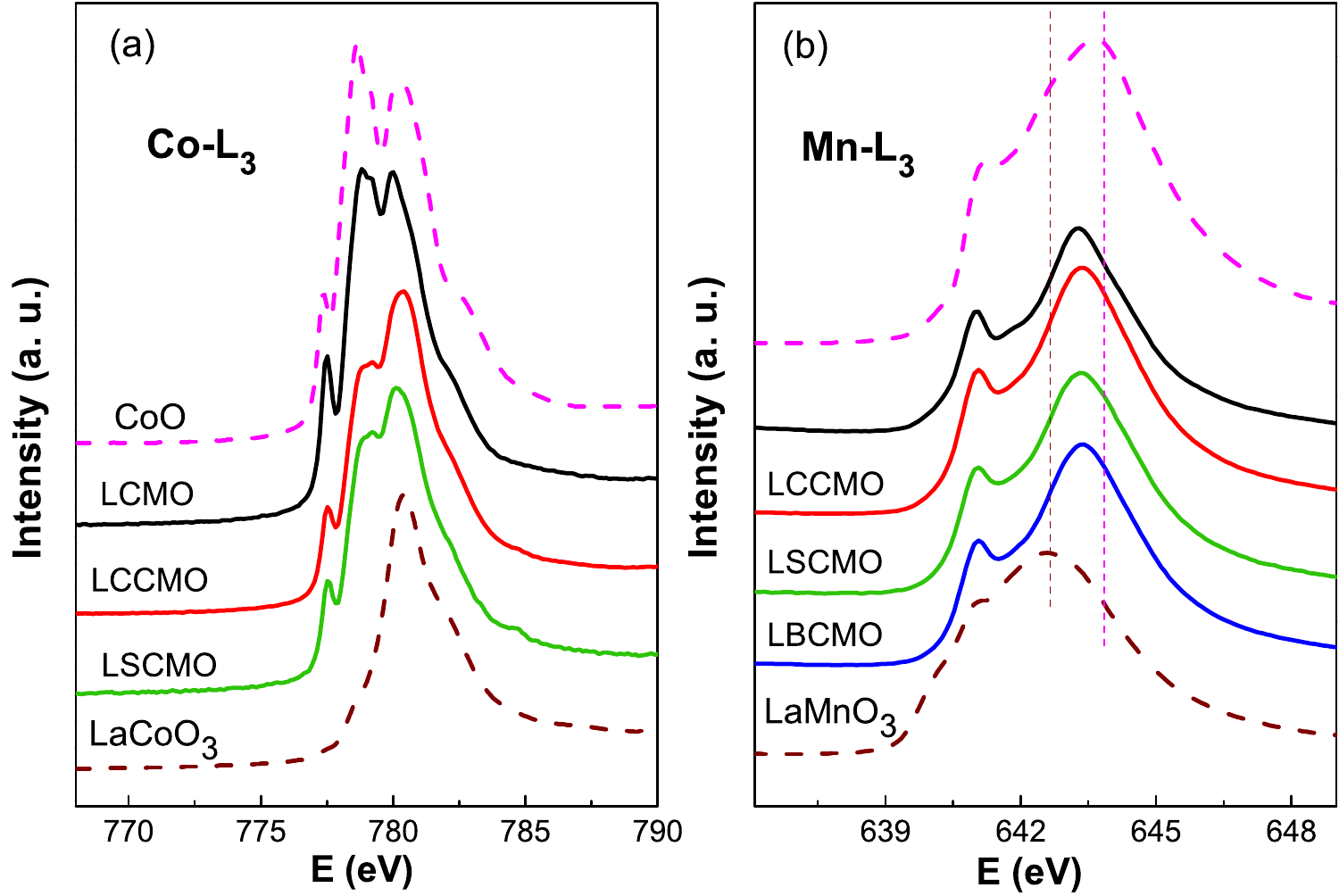}
\end{center}
\caption{(a) Co-$L_3$ and (b) Mn-$L_3$ spectra of La$_{2-x}$A$_{x}$CoMnO$_{6}$ samples at 300 K. The spectra of CoO, LaCoO$_3$, LaMnO$_3$ and CaMnO$_3$ are also displayed as references for Co$^{2+}$, Co$^{3+}$, Mn$^{3+}$ and Mn$^{4+}$, respectively.}
\label{Fig_XAS_L}
\end{figure}

Fig. \ref{Fig_XAS_L}(a) shows the Co-$L_{2,3}$ edge curves of all investigated compounds. Comparing the spectra of LCMO and CoO, it can be seen that the spectral features are quite similar at the low-energy side of the $L_3$ white line, indicating a large concentration of Co$^{2+}$ in this sample. However, at the high-energy side the spectral weight of LCMO is increased in relation to that of CoO, clearly indicating that Co$^{3+}$ is also present in this compound. The spectral feature of LCMO is very similar to those observed for this compound in previous works, with the presence of both Co$^{2+}$ and Co$^{3+}$ valence states \cite{Burnus,Mir}.

The Ba$^{2+}$/Ca$^{2+}$/Sr$^{2+}$ to La$^{3+}$ partial substitution is expected to lead to the increase of Co and/or Mn mean valence, in order to fulfill charge neutrality. For the Ca- and Sr-based samples, Fig. \ref{Fig_XAS_L}(a) shows the increase of the peak about 780 eV, indicative of the increase of the amount of Co$^{3+}$. For LBCMO it was not possible to precisely determine such changes due to the proximity between Co-$L_{2,3}$ and Ba-$M_{4,5}$ edges (see SM \cite{SM}). 

In order to further investigate whether the increase of Co$^{3+}$ concentration in the doped samples is also accompanied by changes in the Mn valence, we carried out XAS spectra around the Mn-$L_{2,3}$ edges. Fig. \ref{Fig_XAS_L}(b) shows the Mn-$L_{3}$ spectra of all investigated samples. The main peak of the curves of La$_{2-x}$A$_{x}$CoMnO$_{6}$ samples lies in between those of the reference spectra for Mn$^{3+}$ and Mn$^{4+}$, indicating that Mn is also in mixed valence state in all investigated samples. Moreover, the spectral features of all samples are very similar, indicating that any doping-induced Mn valence change may be subtle. 

In order to get quantitative estimates of the Co and Mn formal valences we compared  the position of the``center of gravity" of each sample's Co- and Mn-$L_3$ white lines with those of the reference spectra for Co$^{2+}$, Co$^{3+}$, Mn$^{3+}$ and Mn$^{4+}$. Assuming a linear variation from Co$^{2+}$ to Co$^{3+}$ in these systems, we obtained approximately 2.1+, 2.4+ and 2.5+ for the average valences of LCMO, LSCMO and LCCMO, respectively. The same procedure for the Mn-$L_3$ white lines resulted in approximately 3.8+ for LCMO and 3.9+ for all doped samples. These rough estimations indicate that the charge compensation caused by A$^{2+}$ doping at La$^{3+}$ site is manifested mainly at Co valence. 

Several studies have revealed that Co$^{3+}$ is present in non-magnetic LS configuration in  La$_{2-x}$A$_{x}$CoMnO$_{6}$ systems  \cite{Dass,Burnus,Mir}. In the case of the samples investigated here, the doping-induced increase of Mn$^{4+}$ and LS Co$^{3+}$ is supported by the $K$ edge results (see SM \cite{SM}) and by the decrease of $\mu_{eff}$ and $m_{sat}$ observed in the $M(T)$ and $M(H)$ measurements. These changes are expected to remarkably affect the materials magnetic anisotropy and, consequently, the EB effect. However, it must be noted that these rough estimates of the Co and Mn valences would lead to OV in all samples. Assuming the complete stoichiometry of La, Ca/Sr, Co and Mn in the compounds, one would get $\delta$ = 0.1 for LSCMO and 0.05 for LCMO and LCCMO. The OV is known to affect the structure and the magnetization of perovskite compounds \cite{Vasala}, and thus the presence of oxygen holes may directly impact the EB in La$_{2-x}$A$_{x}$CoMnO$_{6}$.

\begin{figure*}
\begin{center}
\includegraphics[width= \textwidth]{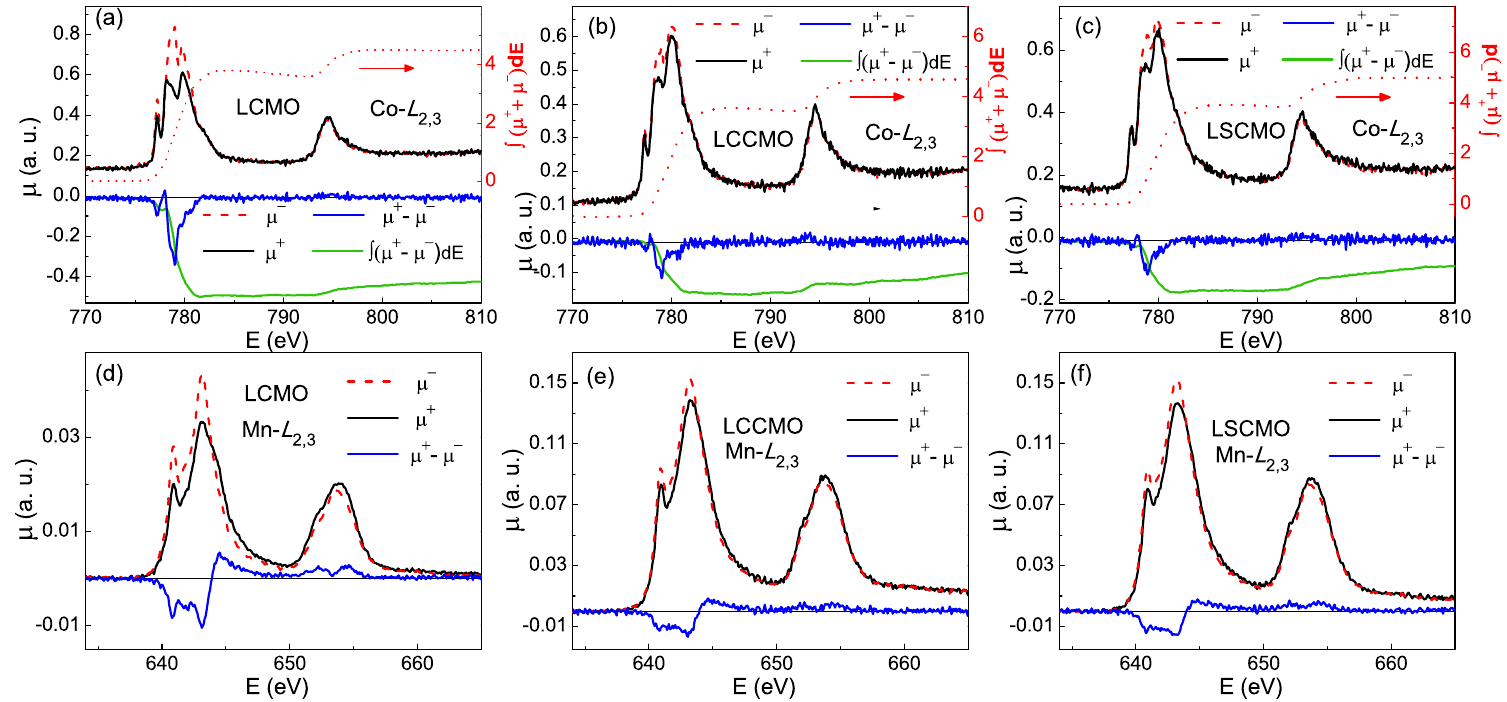}
\end{center}
\caption{(Color online) Co-$L_{2,3}$ spectra of (a) LCMO, (b) LCCMO, (c) LSCMO, and Mn-$L_{2,3}$ spectra of (d) LCMO, (e) LCCMO and (f) LSCMO, taken with circularly polarized x-rays at 14 K. The photon spin was aligned parallel ($\mu^+$, black solid) and antiparallel ($\mu^-$, red dashed) to the 40 kOe magnetic field, and the difference spectra are shown in blue.}
\label{Fig_XMCD}
\end{figure*}

Having established the valences of the Co and Mn ions, we now focus our attention on their
magnetic properties. We carried out XMCD measurements at Co- and Mn-$L_{2,3}$ edges for all investigated samples, at $T$ = 14 K and $H$ = 40 kOe. Fig. \ref{Fig_XMCD} displays the XMCD spectra of LCMO, LCCMO and LSCMO. The results for LBCMO were omitted, due to the presence of Ba-$M_{4,5}$ edges. The red dashed and black solid curves stand, respectively, to $\mu^+$ and $\mu^-$, \textit{i.e.}, for parallel and antiparallel alignments between the photon spin and the magnetic field. The difference spectra, $\triangle \mu$ = $\mu^+$-$\mu^-$, correspond to the blue lines, and the integral curves of the XMCD and XAS spectra are respectively the green and red lines. 

As can be noted, the XMCD curves are largely negative at Co-$L_3$ edge, but are nearly zero at Co-$L_2$ edge. This is a clear indication of a non-negligible orbital contribution to the Co magnetic moment. Despite the different Co average valences observed for these samples, their XMCD signals are spectrally rather similar, indicating that a significant portion of the extra Co$^{3+}$ emerging in LCCMO and LSCMO doped samples may be in LS configuration since this non-magnetic ion is not expected to alter the XMCD results of the HS Co$^{2+}$ ions remaining in the system. For a quantitative analysis of the Co-$L_{2,3}$ XMCD spectra, we have used the sum rules to derive the orbital and spin contributions to the magnetization  \cite{Thole,Carra}
\begin{equation}
m_{orb}=- \frac{4\int_{L_3+L_2}(\mu^+-\mu^-)d\omega}{3\int_{L_3+L_2}(\mu^++\mu^-)d\omega}N_{h}, \label{Eq9}
\end{equation} 
\begin{equation}
\begin{split}
m_{spin} & = -\frac{6\int_{L_3}(\mu^+-\mu^-)d\omega-4\int_{L_3+L_2}(\mu^+-\mu^-)d\omega}{\int_{L_3+L_2}(\mu_++\mu_-)d\omega}\times \\
& N_{h}\left(1+ \frac{7\langle T_z\rangle}{2\langle S_z \rangle}\right)^{-1}.  \label{Eq10}\\
\end{split}
\end{equation} 
where $m_{orb}$ and $m_{spin}$ are the angular and spin magnetic moments in units of $\mu_B$/atom, $L_z$ and $S_z$ denote the projections along $z$ of the angular and spin magnetic momenta, respectively, $N_h$ represent the number of empty 3$d$ states and $T_z$ denotes the magnetic dipole moment. For the $N_h$ we used here as an approximation atomistic values that correspond to 3 holes at the 3$d$ level of Co$^{2+}$. Moreover, $T_z$ has been estimated to be negligible against $S_z$ for TM ions in octahedral symmetry \cite{Teramura,Groot}. With these considerations, Eqs. \ref{Eq9} and \ref{Eq10} become
\begin{equation}
 m_{orb}= \frac{4Q}{R}; \qquad m_{spin}= \frac{18P-12Q}{R},\label{Eq11}
\end{equation}
where $P$, $Q$, and $R$ represent the integrals $\int_{L_3}(\mu^+ - \mu^-)d\omega$, $\int_{L_3+L_2}(\mu^+ - \mu^-)d\omega$ and $\int_{L_3+L_2}(\mu^+ + \mu^-)d\omega$. Nonetheless, a correction to the spin
sum rule must be used for the Co spin moments associated with the relatively weak spin-orbit coupling in the Co 2$p$ core holes. Here the Co$^{2+}$ spin moments extracted from the sum rule were divided by 0.921 to correct the deviation due to the core-hole Coulomb interaction \cite{Teramura}. In addition, corrections for the 80\% of circular polarization must be also included in the calculations. The values so determined are listed in Table \ref{Txmcd}.

\begin{table}

\caption{Results obtained from the Co$-L_{2,3}$ XMCD.}

\label{Txmcd}

\centering

\begin{tabularx}{\columnwidth}{>{\centering}X|>{\centering}X>{\centering}X>{\centering}X>{\centering}X>{\centering}X}

\hline

 & P & Q & R & $m_{orb}(\mu_B)$ & $m_{spin}(\mu_B)$ \tabularnewline  \hline

LCMO & 0.48 & 0.44 & 4.49 & 0.49 & 1.02 \tabularnewline \hline

LCCMO &0.15 &0.12 & 4.56 & 0.16 & 0.38 \tabularnewline \hline

LSCMO & 0.16& 0.12& 5.00 & 0.12 & 0.39 \tabularnewline \hline

\end{tabularx}

\end{table}

Although one can note from Table \ref{Txmcd} a clear decrease of $m_{orb}$ and $m_{spin}$ of the doped samples in relation to LCMO, the values here obtained are smaller than those usually found for Co$^{2+}$ \cite{Burnus} and thus care must be taken in the interpretation of these data. One of the possible reasons for the deviations from the correct values relies in the fact that the estimates of $m_{orb}$ and $m_{spin}$ depend on the value of $R$. Even though the presence of non-magnetic LS Co$^{3+}$ is not expected to alter the XMCD signal of HS Co$^{2+}$, it may contribute to the XAS signal. Moreover, it was used $N_h$ = 3 for all samples, \textit{i.e.}, we assumed that all Co$^{3+}$ are in LS configuration, but it is not possible to know a priori if at least a part of these ions is in HS or even IS state. Finally, it is also possible that the $H$ = 40 kOe used was not enough to achieve a complete magnetization of Co due to the strong magnetocrystalline anisotropy of these polycrystalline materials.

An interesting way to avoid the possible deviations caused by $R$ and $N_h$ is to compute the $m_{orb}$/$m_{spin}$ ratio. For LCMO $m_{orb}$/$m_{spin}$ = 0.48, being in good agreement with previous results found for this compound \cite{Burnus,Mir}. This result is a direct indication of the presence of HS Co$^{2+}$, whose orbital magnetic moment may play an important role on the system magnetic and structural properties. For LCCMO and LSCMO doped samples the $m_{orb}$/$m_{spin}$ ratio is reduced to 0.42 and 0.31, respectively, which could mean the appearance of at least a small fraction of Co$^{3+}$ ions in HS state, since the transition from HS Co$^{2+}$ ($t_{2g}^{5}e_g^{2}$, S = 3/2, $\tilde{L}$ = 1) to HS Co$^{3+}$ ($t_{2g}^{4}e_g^{2}$, $S$ = 3/2, pseudo-orbital moment $\tilde{L}$ = 1) leads to the increase of the spin moment \cite{Hollmann}. However, the presence IS state, already proposed for Sr-doped LCMO compounds \cite{Taraphder}, can not be completely ruled out in spite the expected decrease of the spin moment of IS Co$^{3+}$ ($t_{2g}^{5}e_g^{1}$, S = 1) in relation to HS Co$^{2+}$. The reduction of the orbital contribution may be even greater due to the ordering of the split $e_g$ orbitals and to their strong hybridization with the O-2$p$ orbitals. The decrease of the Co moment observed in the XMCD, as well as for the net moment observed in the macroscopic magnetization results, indicate that the occurrence of IS Co$^{3+}$, if so, may be present in a minor part of Co ions, possibly in the proximity to Mn$^{3+}$ ions, which would result in local Jahn-Teller deformations \cite{Dass}. The Co$^{3+}$ HS/IS state conjectures are speculative, and other mechanisms may play a role on the decrease of $m_{orb}$/$m_{spin}$ observed in the doped samples. The OV, for instance, may remarkably affect the Co magnetic moment \cite{Vasala,Miao}. 

Fig. \ref{Fig_XMCD} also displays the XMCD spectra at Mn-$L_{2,3}$ edges. In general, the sum rules analysis is not fully valid for systems containing small 2$p$ core-hole spin orbit coupling such as Mn, for which the error in the calculated effective spin moment is very large \cite{Groot}. Nevertheless, it is important to note that the XMCD spectra are negative at both Mn- and Co-$L_3$ edges, confirming that the FM coupling between Co and Mn is present in all samples.

\begin{figure}
\begin{center}
\includegraphics[width=0.47 \textwidth]{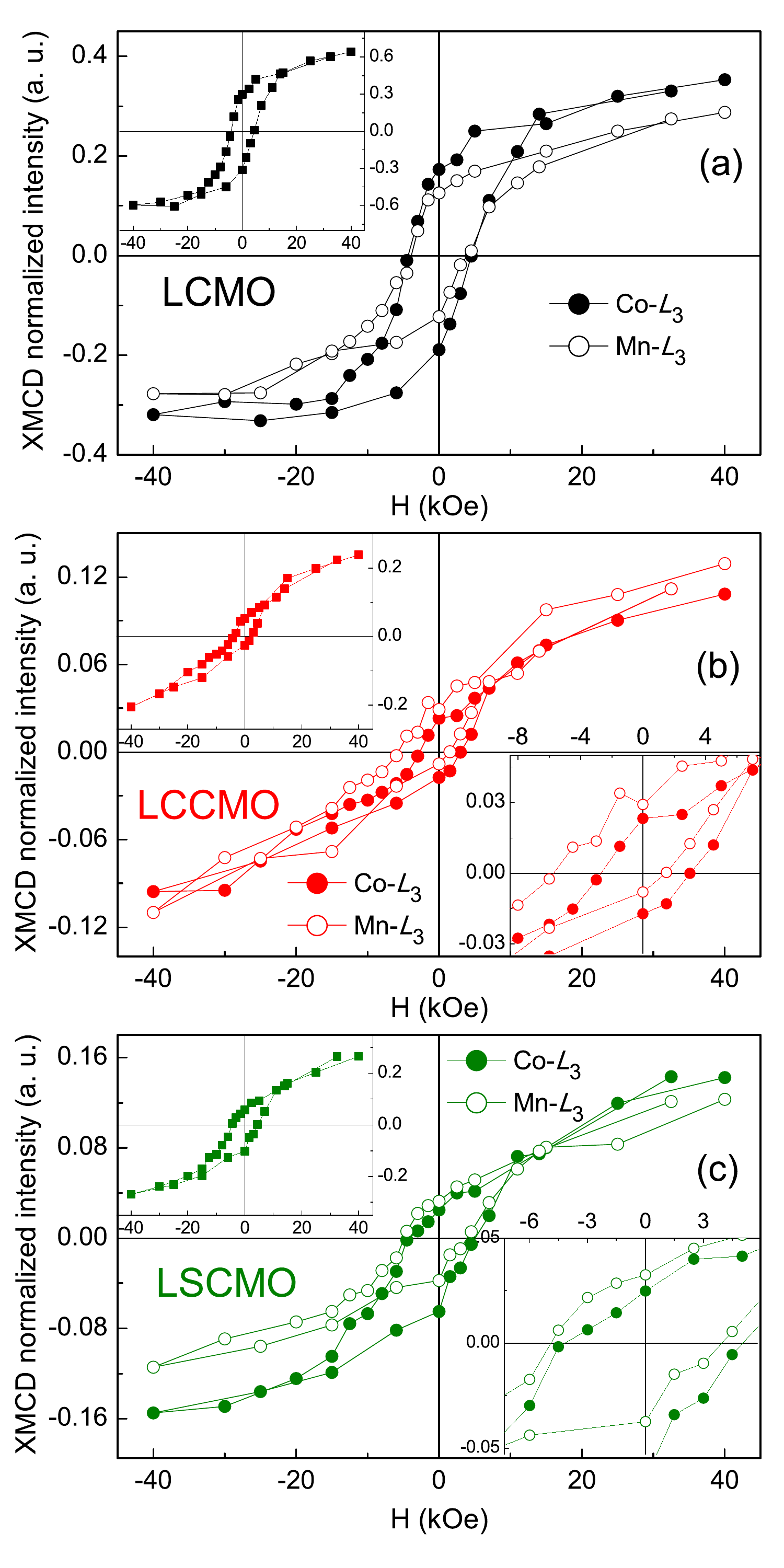}
\end{center}
\caption{XMCD hysteresis loops of Co and Mn $L_3$ edges for (a) LCMO, (b) LCCMO and (c) LSCMO, measured at 14 K. The upper insets display the curves resulting from the sum of Co and Mn signals, and the bottom ones show magnified views of the loops close to the coercive field regions.}
\label{Fig_XMCDloop}
\end{figure}

To get a deeper insight about the role played by Co and Mn on the ZEB effect, we carried
element-specific hysteresis loops in all samples by monitoring the Co- and Mn-$L_3$ edge XMCD signals as a function of applied magnetic field. Fig. \ref{Fig_XMCDloop}(a) shows the hysteresis loops for Co (solid circles) and Mn (open circled) in LCMO, for which the XMCD signals are normalized by their isotropic XAS signals at the $L_3$ peak. The data were taken at $T$ = 14 K and using $H_{max}$ = 40 kOe, after ZFC the sample. As expected for this non-ZEB material, there is no shift of the Co or Mn curves along $H$ axis.  The inset shows the loop resulting from the sum of Co and Mn signals, which is very similar to the $M(H)$ curve obtained in Fig. \ref{Fig_MxH}(a).

Very interesting results were found for the doped samples. Fig. \ref{Fig_XMCDloop}(b) displays the hysteresis loops obtained for LCCMO, where it can be seen that the curve for Mn is shifted to the left, while the one for Co is shifted to the right. It can also be noted an overall decrease of the XMCD signal in relation to LCMO parent compound, in resemblance to the results found in the macroscopic $M(H)$ curves. The fact the reduction is not as large as that observed in the macroscopic magnetization results is probably related to the different $T$ and $H$ at which the investigations were performed. These results are indicative of the emergence of AFM coupling, with the decrease of the Co signal in relation to Mn being caused by the increased amount of LS Co$^{3+}$. The uncompensated shifts of Co and Mn curves result in the ZEB effect, as already observed in the macroscopic $M(H)$ curve. A similar result was found for LSCMO, as can be seen in Fig. \ref{Fig_XMCDloop}(c). The XMCD loops obtained for LBCMO were omitted in this figure due to the proximity between Ba-$M_{5}$ and Co-$L_3$ edges. Nevertheless, the Mn-$L_3$ XMCD loop of LBCMO is shifted to the left, in similarity to LCCMO and LSCMO (see SM \cite{SM}). We are aware of the importance of conducting Ligand-Field Multiplet calculations to quantitatively describe our results. Such analysis is being conducted and will be published elsewhere.

The ZEB effect here observed for the doped samples can be understood in terms of the AFM coupling between Co and Mn in the low-$H$ regions of the $M(H)$ curves, \textit{i.e.}, the regions encompassing the positive and negative coercive fields. During the $M(H)$ cycle, the initial magnetization process induces the alignment of Mn ions toward $H$ (positive) direction. A part of these ions are AFM coupled to Co ions and, during the $H$ cycling, some of them become pinned toward the positive field direction while some Co ions are consequently pinned into negative direction. Since Mn$^{3+}$/Mn$^{4+}$ magnetic moment is larger than that of LS Co$^{3+}$, this AFM coupling is uncompensated, leading to the shift of the $M(H)$ curve toward the left side of $H$ axis. As the proportion of LS Co$^{3+}$ ions increases, the uncompensation increases, which corroborates the largest $H_{EB}$ observed for LSCMO. The XMCD results indicated a smaller Co magnetic moment for this compound. However, the possibility that the changes in the ZEB effect are related to OV cannot be excluded, since LSCMO also presents the largest $\delta$. 

It is also possible that the SG-like behavior observed for the doped compounds comes from the Mn ions surrounded by non-magnetic LS Co$^{3+}$, since they would be under the effect of weaker magnetic interactions, which may also lead to frustration. Increasing the amount of LS Co$^{3+}$ ions would result in the observed reduction of the systems overall magnetization, but it could lead to the increase of the amount of SG-like phase.

\section{Summary}

In summary, in this work we thoroughly investigated the structural, electronic and magnetic properties of LCMO, LBCMO, LCCMO and LSCMO compounds. Our SXRD results indicate more symmetrical crystal structures for Ba- and Sr-doped samples. It was also found that all samples present cationic disorder of Co and Mn, and that the doped compounds present a small amount of phase segregation. These results are corroborated by the Raman spectroscopy measurements, and are believed to be related to the larger ZEB effect observed for LBCMO and LSCMO. 

In addition to the two FM transitions observed for LCMO, for the doped samples it was found the emergence of a third anomaly at low-$T$ in the $M(T)$ measurements that is related to the formation of a CG behavior, as concluded from AC magnetic susceptibility and $\mu$SR experiments. This third magnetic transition is most likely related to the Co$^{3+}$--O--Mn$^{4+}$ AFM coupling. XAS measurements indicate mixed valence states Co$^{2+}$/Co$^{3+}$ and Mn$^{4+}$/Mn$^{3+}$ in all samples, and also that Ba$^{2+}$/Ca$^{2+}$/Sr$^{2+}$ partial substitution at La$^{3+}$ site leads to a large increase of Co average valence, with subtle changes of the Mn formal valence. The XAS data also suggest the presence of OV in the samples, which may also play an important role on their magnetic properties. Our XMCD results indicate that the ZEB effect observed for the doped samples is related to uncompensated AFM coupling between Co and Mn. The reduction of Co magnetic moment observed for LSCMO, induced by the increased proportion of Co$^{3+}$ and/or OV, augments this uncompensation and results in its large ZEB effect. Similar result may be found for LBCMO.

\begin{acknowledgements}
This work was supported by Conselho Nacional de Desenvolvimento Cient\'{i}fico e Tecnol\'{o}gico (CNPq) [No. 400134/2016-0], Funda\c{c}\~{a}o Carlos Chagas Filho de Amparo \`{a} Pesquisa do Estado do Rio de Janeiro (FAPERJ), Funda\c{c}\~{a}o de Amparo \`{a} Pesquisa do Estado de Goi\'{a}s (FAPEG), Funda\c{c}\~{a}o de Amparo \`{a} Pesquisa do Estado de S\~{a}o Paulo (FAPESP) and Coordena\c{c}\~{a}o de Aperfei\c{c}oamento de Pessoal de N\'{i}vel Superior (CAPES).  The authors thank the DXAS, PGM, XDS and XPD staffs of LNLS for technical support and LNLS for the concession of beam time (proposals No. 20160546, 20160578, 20170057, 20170310, 20170612 and 20180745). E.S., F.J.L. and E.B.S. acknowledge support by a joint DFG-FAPERJ project Li 244/12. F.J.L. is grateful for a fellowship by FAPERJ.
\end{acknowledgements}

\section{Supplementary Material}

\subsection{S1: Synthesis and crystal structures}

La$_{2}$CoMnO$_{6}$ (LCMO), La$_{1.5}$Ca$_{0.5}$CoMnO$_{6}$ (LCCMO), La$_{1.5}$Sr$_{0.5}$CoMnO$_{6}$ (LSCMO) and La$_{1.5}$Ba$_{0.5}$CoMnO$_{6}$ (LBCMO) polycrystalline samples were synthesized by conventional solid state reaction, in air atmosphere and ambient pressure. Stoichiometric amounts of La$_{2}$O$_{3}$, CaO/SrCO$_{3}$/BaCO$_{3}$, MnO and Co$_{3}$O$_{4}$ were mixed and sent to the furnace with different heat treatments, depending on the compound. To grow LCMO the reagents were first heated at $1000^{\circ}$C for 24 hours. Subsequently, the material was ground and sent to a second furnace step of 12 hours at $1300^{\circ}$C. Finally, the powder was re-ground, pressed into pellet and heated at $1300^{\circ}$C for 12 hours. LCCMO was obtained after three heat treatments. At first the precursor oxides were sent to $800^{\circ}$C for 12 hours. Then the material was re-ground and sent to $1200^{\circ}$C for 24 hours. At last it was ground, pressed into pellet and sent to $1400^{\circ}$C for 24 hours. To produce LSCMO the precursor powders were sent to a first heating step of $750^{\circ}$C for 10 hours. Then the material was sent to $1400^{\circ}$C for 24 hours, with intermediate grinding, and then the powder was re-ground, pressed into pellet and sent again to $1400^{\circ}$C for 12 hours. Finally, LBCMO was obtained after three heating treatments. At first the reagents were mixed and sent to $850^{\circ}$C for 10 hours. After the first treatment the material was re-ground and sent to a furnace step of $1300^{\circ}$C for 10 hours. Then the powder was pressed into pellet and sent again to $1300^{\circ}$C for 10 hours.

The formation of each desired phase was initially checked using laboratory X-ray powder diffraction measurements, performed at room temperature ($T$) in a Bruker \textit{D8 Discover} diffractometer, coupled with a Johanson monochromator and operating with Cu $K_{\alpha}$ radiation. After the formation of the desired compound was confirmed, the crystal structure of each sample was investigated in detail using Synchrotron XRD (SXRD) measurements, performed at several $T$ at the XPD beamline of the Brazilian Synchrotron Light Laboratory (LNLS) using Bragg-Brentano geometry. 

\begin{figure}
\begin{center}
\includegraphics[width=0.48 \textwidth]{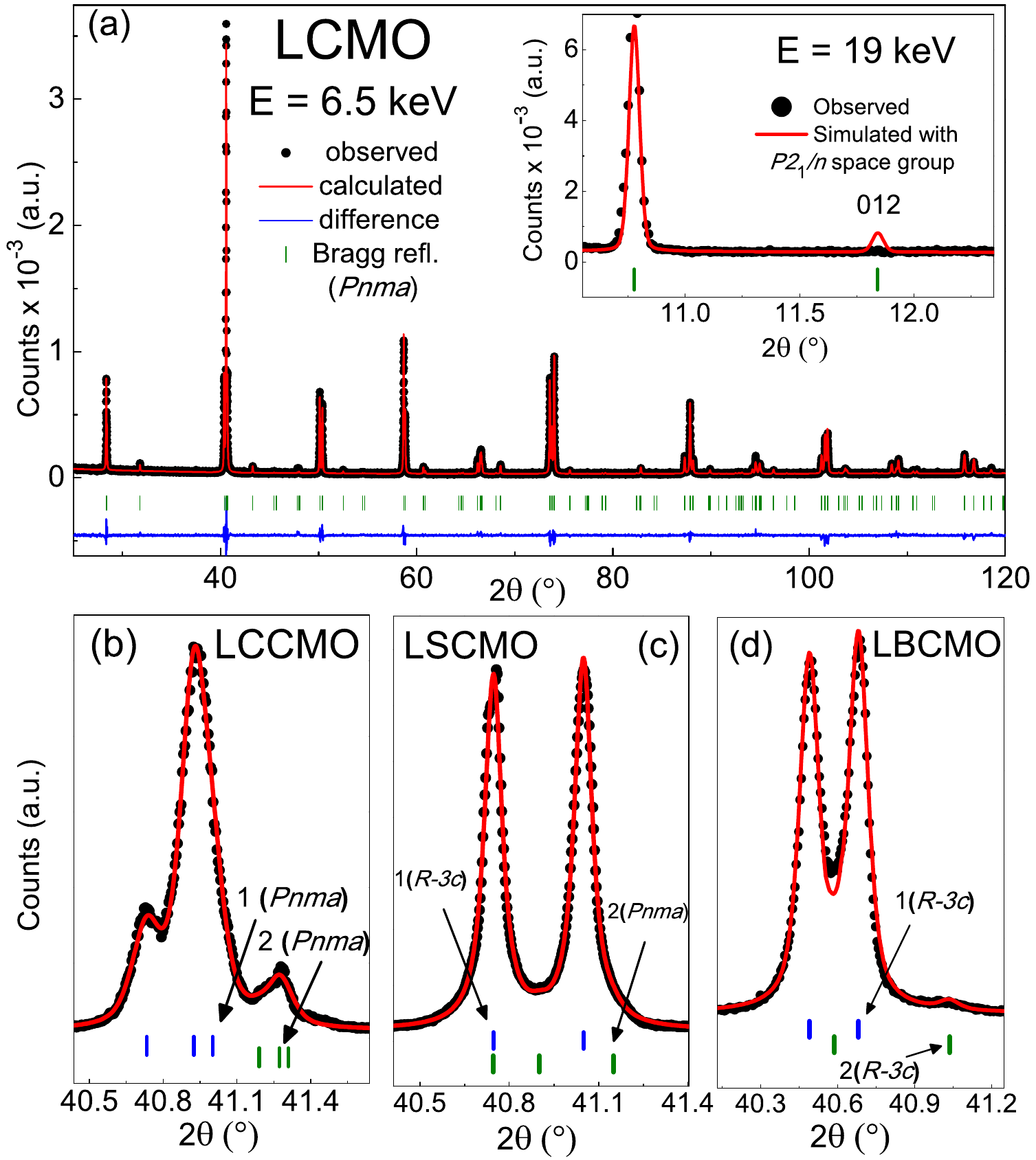}
\end{center}
\caption{(a) Room-$T$ SXRD of LCMO measured with $\lambda$ = 1.9074 \AA. The inset shows a magnified view of the $E$ = 19 keV SXRD in the region where the 012 reflection of the $P2_{1}/n$ space group is expected. The solid line represent a SXRD curve simulated in $P2_{1}/n$ space group. (b), (c) and (d) show magnified views of the main peaks of $E$ = 6.5 keV SXRD data of LCCMO, LSCMO and LBCMO, respectively, highlighting the presence of two phases in the doped samples. The blue and dark grey vertical bars correspond to the Bragg reflections for the main phase and for the second phase, respectively.}
\label{FigSXRD_s}
\end{figure} 

There is a debate in the literature about Co and Mn being ordered or not along the lattice in LCMO compound. To check that, we performed a long-lasting measurement using an incident wavelength of photon energy $E$ = 6.5 keV ($\lambda$ = 1.9074 \AA), which maximize the difference between Co and Mn scattering factors. Fig. \ref{FigSXRD_s}(a) displays the SXRD collected for LCMO sample, for which it was not observed the Bragg peaks associated to the ordered $P2_{1}/n$ monoclinic space group. This SXRD pattern could be successfully refined in the disordered $Pnma$ space group. To confirm the absence of such Bragg reflections, we also performed SXRD using an incident wavelength of photon energy $E$ = 19 keV ($\lambda$ = 0.6525 \AA), which maximize the intensities of the peaks. The inset of Fig. \ref{FigSXRD_s}(a) shows a magnified view of the resulting diffractogram  in the region where the 012 reflection of $P2_{1}/n$ space group should be found. It also displays a curve simulated using the instrumental parameters and background function extracted from the Rietveld refinement. As can be seen, the expected peak is not observed,thus confirming a disordered arrangement of Co and Mn in our LCMO sample. 

We also performed SXRD measurements whit $\lambda$ = 1.9074 \AA in LCCMO, LSCMO and LBCMO. Fig. \ref{FigSXRD_s}(b)-(d) show magnified views of the main peaks obtained for these samples, where the presence of two phases become clear. The Rietveld refinement of each diffractogram was performed with GSAS+EXPGUI software \cite{GSAS}, using a Shifted Chebyschev function for the background (15 terms) and a pseudo-Voigt function (function 2) for the instrumental profile parameters. The isotropic $T$ factors ($U_{iso}$) were constrained to be the same for the B-site ions, as well as for the A-site and oxygen ions, and for the doped samples the profile parameters were constrained to be the same for the two phases. The main results obtained from the refinement are displayed in Table \ref{TS1}.

\begin{table}
\renewcommand{\arraystretch}{1.2}
\caption{Main results obtained from the Rietveld refinements of the $E$ = 6.5 keV SXRD data. For the doped samples, the lattice parameters displayed refer to the main phase.}
\label{TS1}
\resizebox{\columnwidth}{!}{
\begin{tabular}{c|cccc}
\hline \hline
Sample & LCMO & LCCMO & LSCMO & LBCMO \\
\hline
main phase & $Pnma$ & $Pnma$ (94\%) & $R\bar{3}c$ (91\%) & $R\bar{3}c$ (96\%) \\

secondary phase & - & $Pnma$ (6\%) & $Pnma$ (9\%) & $R\bar{3}c$ (4\%) \\

$a$ (\AA) & 5.4742(1) & 5.4335(1) & 5.4704(1) & 5.5023(1) \\

$b$ (\AA) & 7.7551(1) & 7.6890(1) & 5.4704(1) & 5.5023(1) \\

$c$ (\AA) & 5.5171(1) & 5.4702(1) & 13.2574(1) & 13.3869(1) \\

$R_p$ & 7.9 & 8.6 & 5.3 & 7.3 \\

$R_{wp}$ & 10.9 & 12.2 & 8.7 & 9.7 \\
\hline \hline
\end{tabular}}
\end{table}

The SXRD measurement at lower $T$ were performed with $E$ = 9 keV ($\lambda$ = 1.3776 \AA). For the Rietveld refinements of these results the phase fractions were kept fixed at the values obtained from the $E$ = 6.5 keV results, and the background and profile functions used were the same described above. In order to avoid divergence, and to get a reliably evolution of the structure with $T$, for the low-$T$ refinements we kept the profile parameters fixed at the values obtained from the room-$T$ data. 

\subsection{S2: Magnetization results}

\begin{figure}
\begin{center}
\includegraphics[width=0.5 \textwidth]{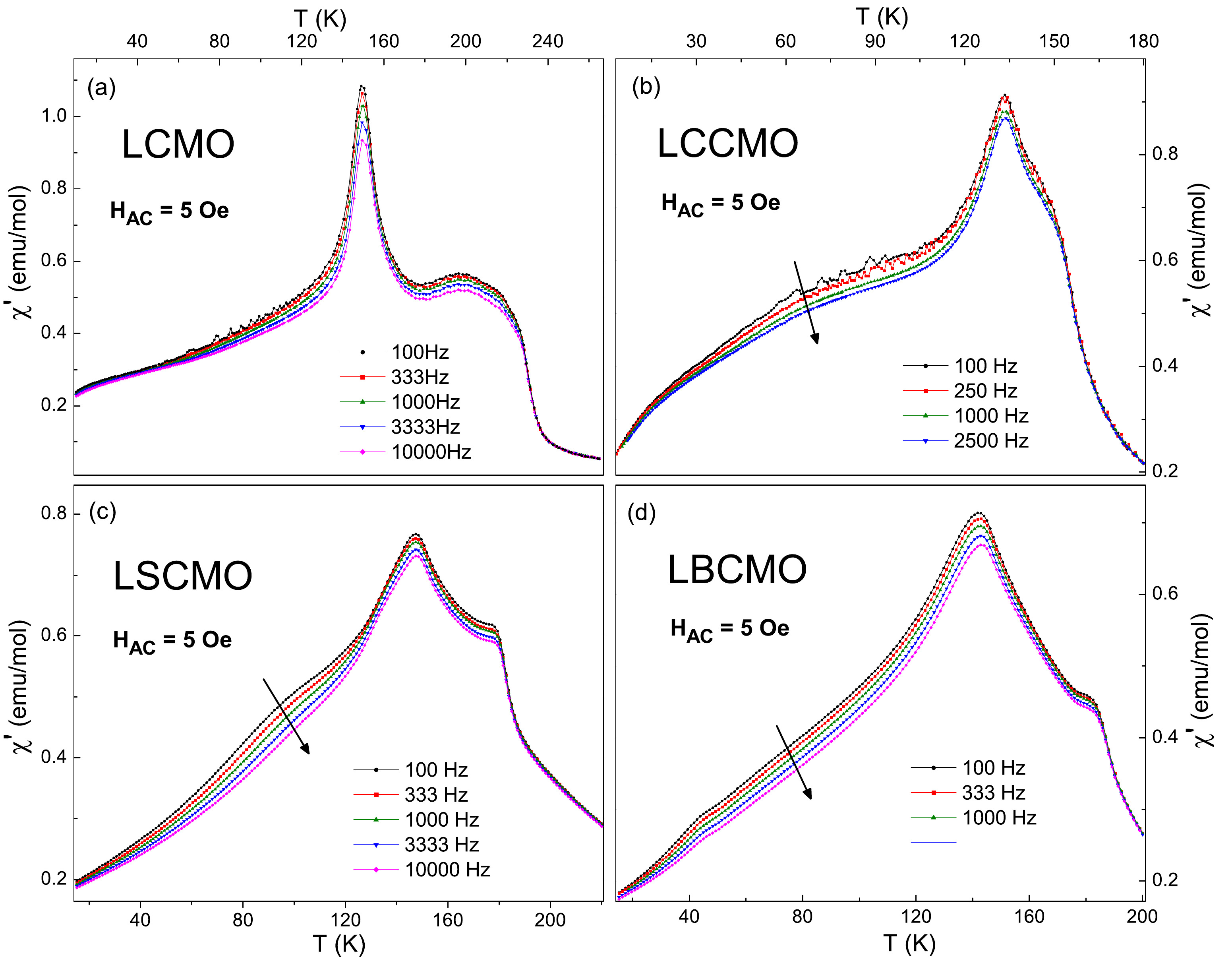}
\end{center}
\caption{$\chi'$ as a function of $T$ for (a) LCMO, (b) LCCMO, (c) LSCMO and (d) LBCMO.}
\label{Fig_chi1}
\end{figure}

\begin{figure*}
\begin{center}
\includegraphics[width= \textwidth]{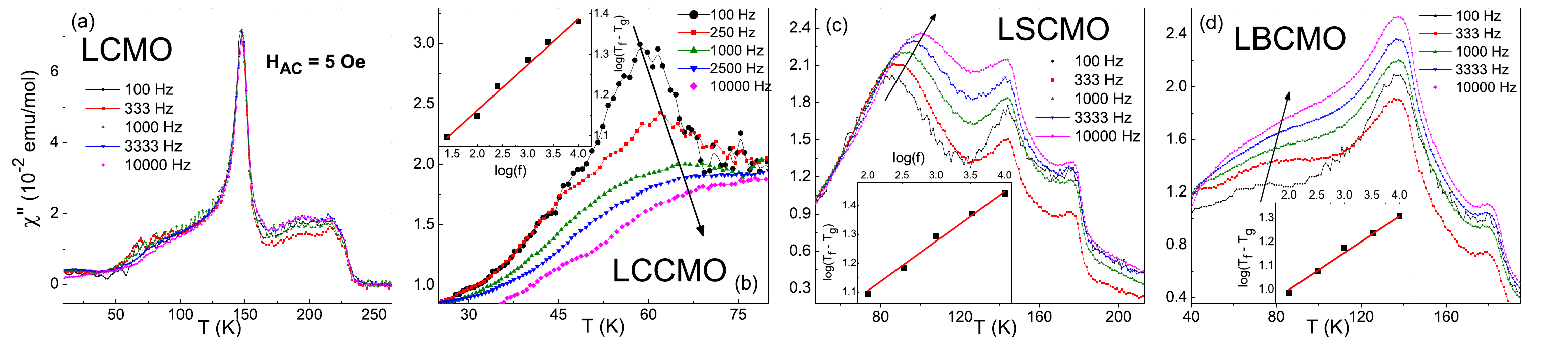}
\end{center}
\caption{$\chi"$ as a function of $T$ for (a) LCMO, (b) LCCMO, (c) LSCMO and (d) LBCMO. The insets show $T_f$ as a function of $f$, where the solid lines represent the best fits with the power law equation of the dynamical scaling theory.}
\label{Fig_chi2}
\end{figure*}

AC magnetic susceptibility ($\chi_{AC}$) as a function of $T$ was measured in the samples of interest at several frequencies ($f$) in the range 25-10000 Hz, using AC magnetic field $H_{AC}$ = 5 Oe. Figs. \ref{Fig_chi1} and \ref{Fig_chi2} show the real ($\chi$') and imaginary ($\chi$") parts of the AC susceptibility for some selected $f$. For LCMO, although one can notice a decrease of the magnitude of the peaks with increasing $f$, there is not a monotonic variation of the peak position with increasing $f$. This suggests that these changes are associated to the FM transitions already observed in the DC results, rather than to spin glass (SG)-like behavior. The decrease in the magnitude of the peak with increasing $f$ is also expected for conventional FM/AFM transitions \cite{Fujiki,Balanda}. Besides, this compound was exhaustively investigated in the last decades and, to the best of our knowledge, SG-like behavior was not suggested.

For the Ca-, Ba- and Sr- doped samples, we observe the same decrease with $f$ of the peaks related to their FM transitions, $T_{C1}$ and $T_{C2}$, as expected. However, we can also note the presence of $f$-dependent peaks characterized by changes on the slope of the curves. This becomes evident for the $\chi"$ curves, Fig. \ref{Fig_chi2}, where cusps are observed at lower $T$. The peak position increases with increasing $f$, suggesting SG-like behavior.

The peak positions extracted from $\chi"$ for the doped samples could be well fitted by the power law equation of the dynamic scaling theory \cite{Mydosh,Souletie}, commonly used to investigate SG-like systems. The fittings can be observed on the insets of Fig. \ref{Fig_chi2}, and the main results obtained from the fits are displayed in the main text.

An important detail to be discussed for LSCMO is that this compound was previously characterized as SG \cite{Murthy,Murthy2}, while our results suggest cluster glass (CG) behavior. The fact is that we have produced our sample by solid state reaction, and the sample reported on Refs. \onlinecite{Murthy,Murthy2} was synthesized using the sol-gel method. As discussed in Ref. \onlinecite{ZEBmodel}, these materials' magnetic properties seems to be intrinsically related to the magnetic interfaces, which are strongly correlated to their morphology. The domain size and morphology of magnetic oxides are known to deeply depend on the method at which the sample was prepared \cite{Serrate,Frey}.

\subsection{S3: Muon spins rotation and relaxation}

Muon spin rotation and relaxation ($\mu$SR) experiments were performed using nearly 100\% spin-polarized positive muons. The measurements were carried in zero field (ZF) and weak transverse field (wTF, field applied perpendicular to the initial muon spin direction) modes. The data were acquired at several $T$ between 5 and 300 K.  wTF experiments were performed under a field of  $H$ = 50 Oe.

The explicit $T$ dependent $\mu$SR asymmetry $A(t)$ in ZF spectra is given by
\begin{equation}
\begin{split}
A(t) & = A(t=0)\{A_{PM1}e^{-\lambda_{1}t} + A_{PM2}e^{-(\lambda_{2}t)^{\beta}} \\
& \cdot A_{int}[\frac{2}{3}cos(\omega_{int}t)e^{-\lambda_{T}t} + \frac{1}{3}e^{-\lambda_{L}t}] \\
& + A_{KT}\{\frac{1}{3} + \frac{2}{3}[1-(\sigma t)^2]\cdot e^{-\frac{(\sigma t)^2}{2}}\}\}. \label{EqS1} \\
\end{split} 
\end{equation}
$A_{PM1}$, $A_{PM2}$, $A_{int}$, and $A_{KT}$ are the partial asymmetry contributions from the slowly damped paramagnetic, the stretched exponentially damped paramagnetic, the internal field and the Gaussian Kubo-Toyabe signals, respectively. $\lambda_1$ and $\lambda_2$ are the damping parameters of the paramagnetic signals, $\lambda_T$ and $\lambda_L$ are the parameters for transverse and longitudinal damping of the internal field depolarization function. $\beta$  is the stretching parameter of the PM2 signal damping, and $\omega_{int}$ is the spontaneous muon spin rotation frequency due to an internal field, $B_{int}$, at the muon site. Fig. \ref{Fig_S_ZF_LBCMO} below shows the $B_{int}$ as a function of $T$ extracted from the ZF $\mu$SR spectra of LBCMO.

\begin{figure}
\begin{center}
\includegraphics[width=0.45 \textwidth]{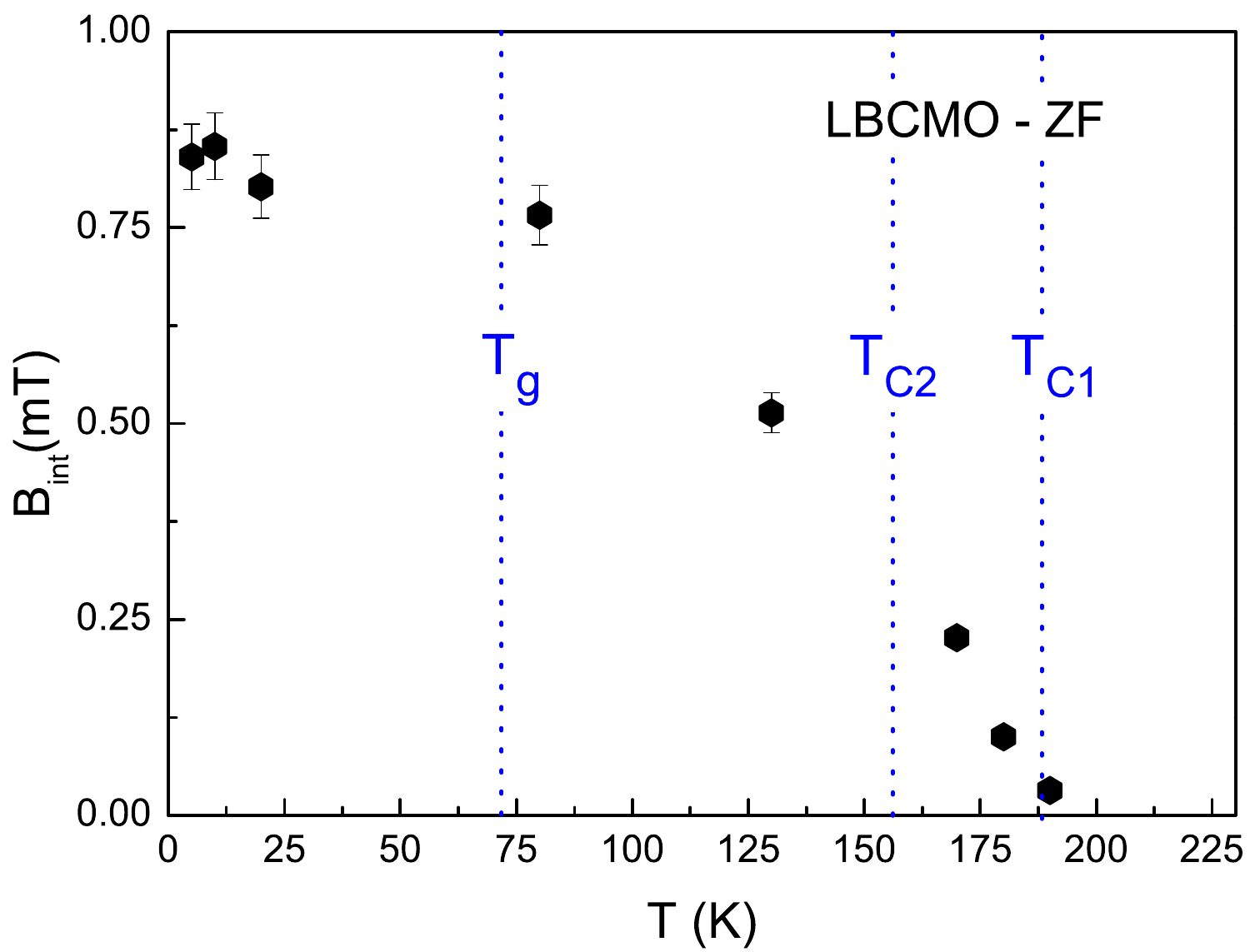}
\end{center}
\caption{$T$ dependence of internal magnetic field at muon site from spontaneous rotations in ZF spectra of LBCMO.}
\label{Fig_S_ZF_LBCMO}
\end{figure}

Overall, the $\mu$SR results found for the LCCMO and LSCMO samples are very similar to those observed for LBCMO. Fig. \ref{Fig_S_ZF_LSCMO} below shows the ZF $mu$SR spectra of LSCMO at some selected temperatures. Comparing the spectra carried at $T$ above and below the first magnetic transition, one can note changes in the initial asymmetries that agree with the results obtained from $M(T)$ and Raman spectroscopy. 

\begin{figure}
\begin{center}
\includegraphics[width=0.46 \textwidth]{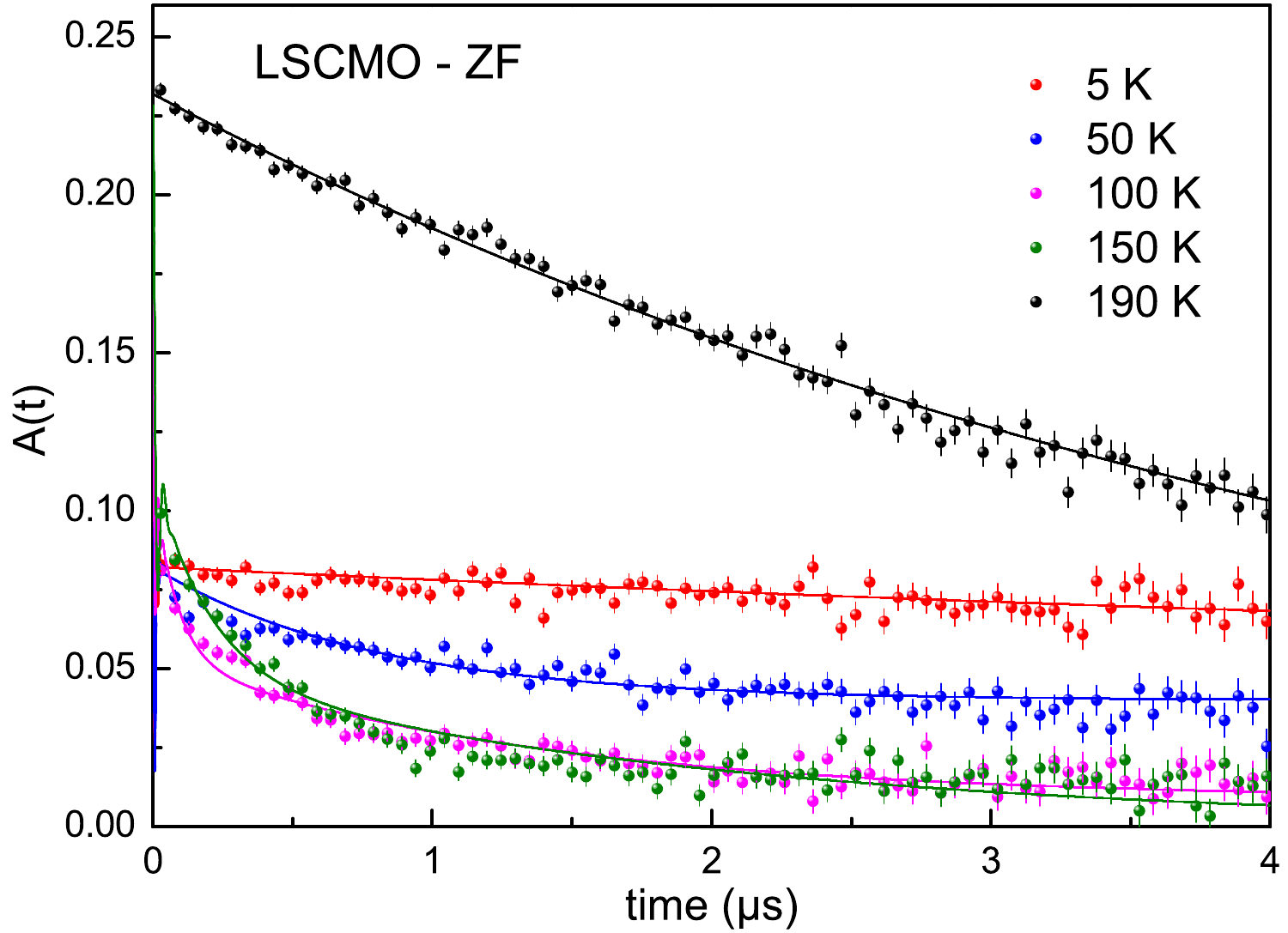}
\end{center}
\caption{ZF $\mu$SR spectra of LSCMO at some selected temperatures.}
\label{Fig_S_ZF_LSCMO}
\end{figure}

\begin{figure}
\begin{center}
\includegraphics[width=0.46 \textwidth]{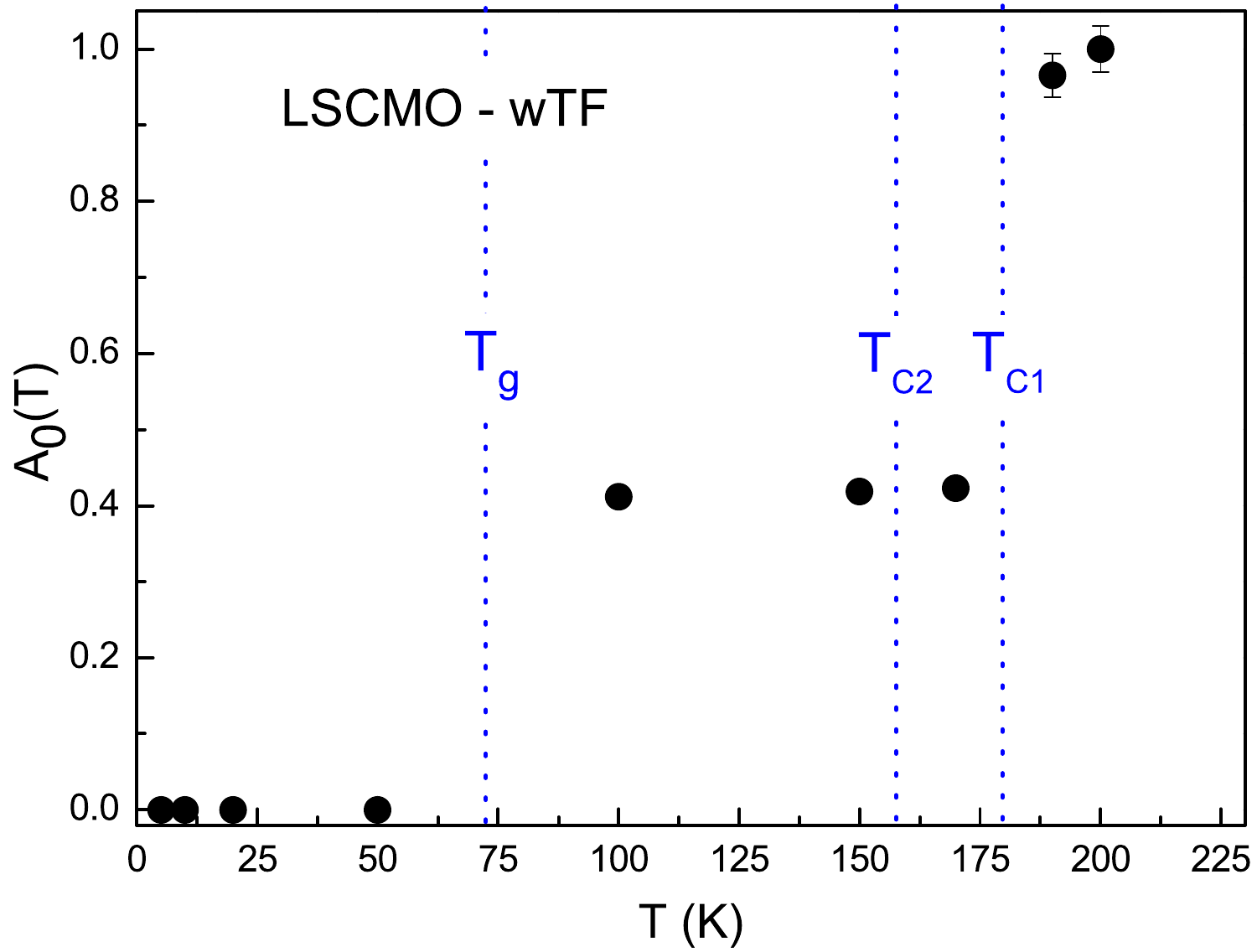}
\end{center}
\caption{Variation of weakly damped paramagnetic fraction with temperature derived from wTF spectra of LSCMO. Indicated marks for $T_{C1}$, $T_{C2}$ and $T_g$ are taken from magnetization and susceptibility data.}
\label{Fig_S_TF_LSCMO}
\end{figure}

Fig. \ref{Fig_S_ZF_LCCMO} below shows the ZF $\mu$SR spectra of LCCMO at some selected $T$s, while Fig. \ref{Fig_S_TF_LCCMO} displays the $T$ dependence of weakly damped paramagnetic fraction in LCCMO, extracted from wTF spectra.

\begin{figure}
\begin{center}
\includegraphics[width=0.46 \textwidth]{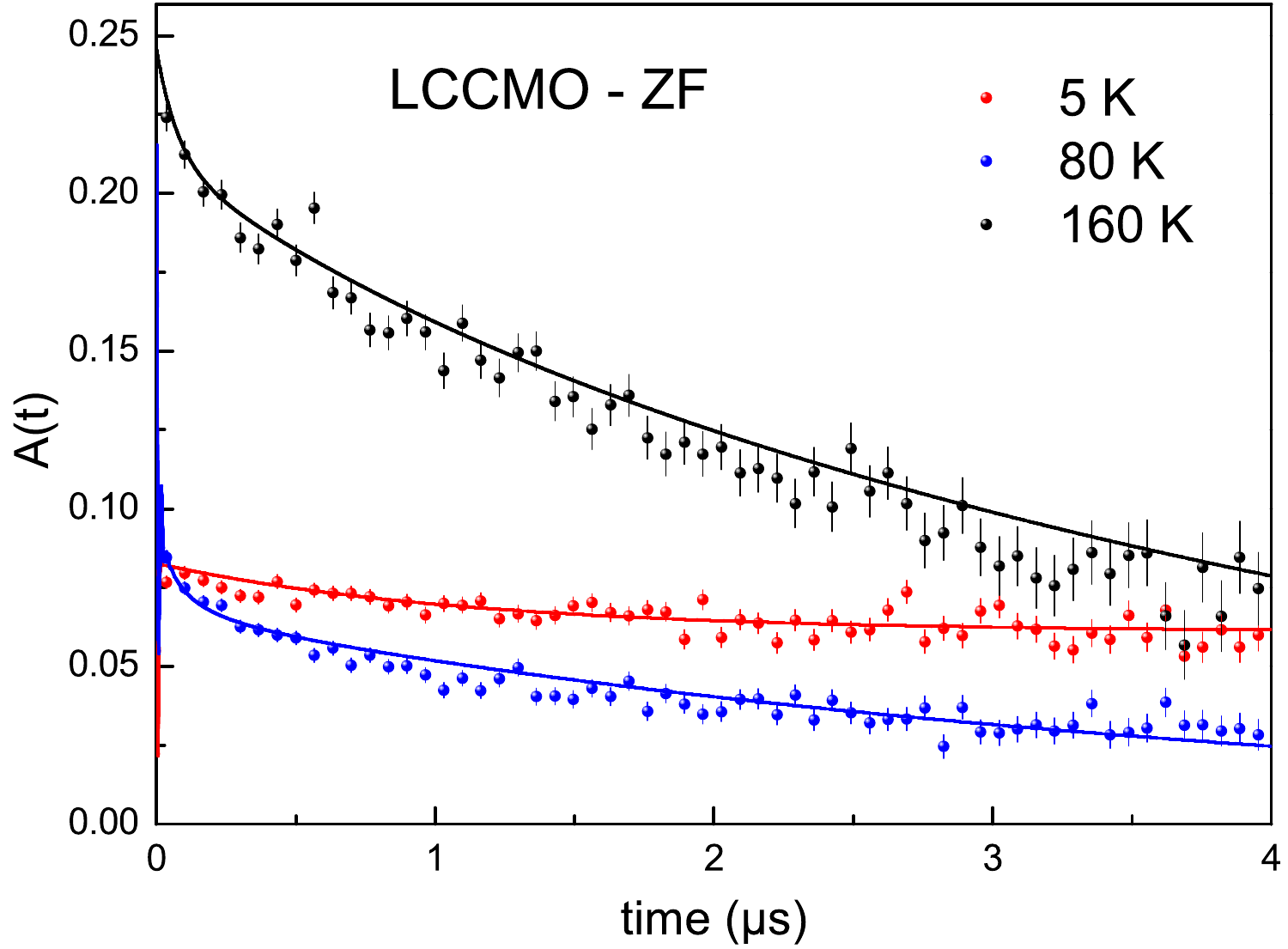}
\end{center}
\caption{ZF $\mu$SR spectra of LCCMO at some selected temperatures.}
\label{Fig_S_ZF_LCCMO}
\end{figure}

\begin{figure}
\begin{center}
\includegraphics[width=0.46 \textwidth]{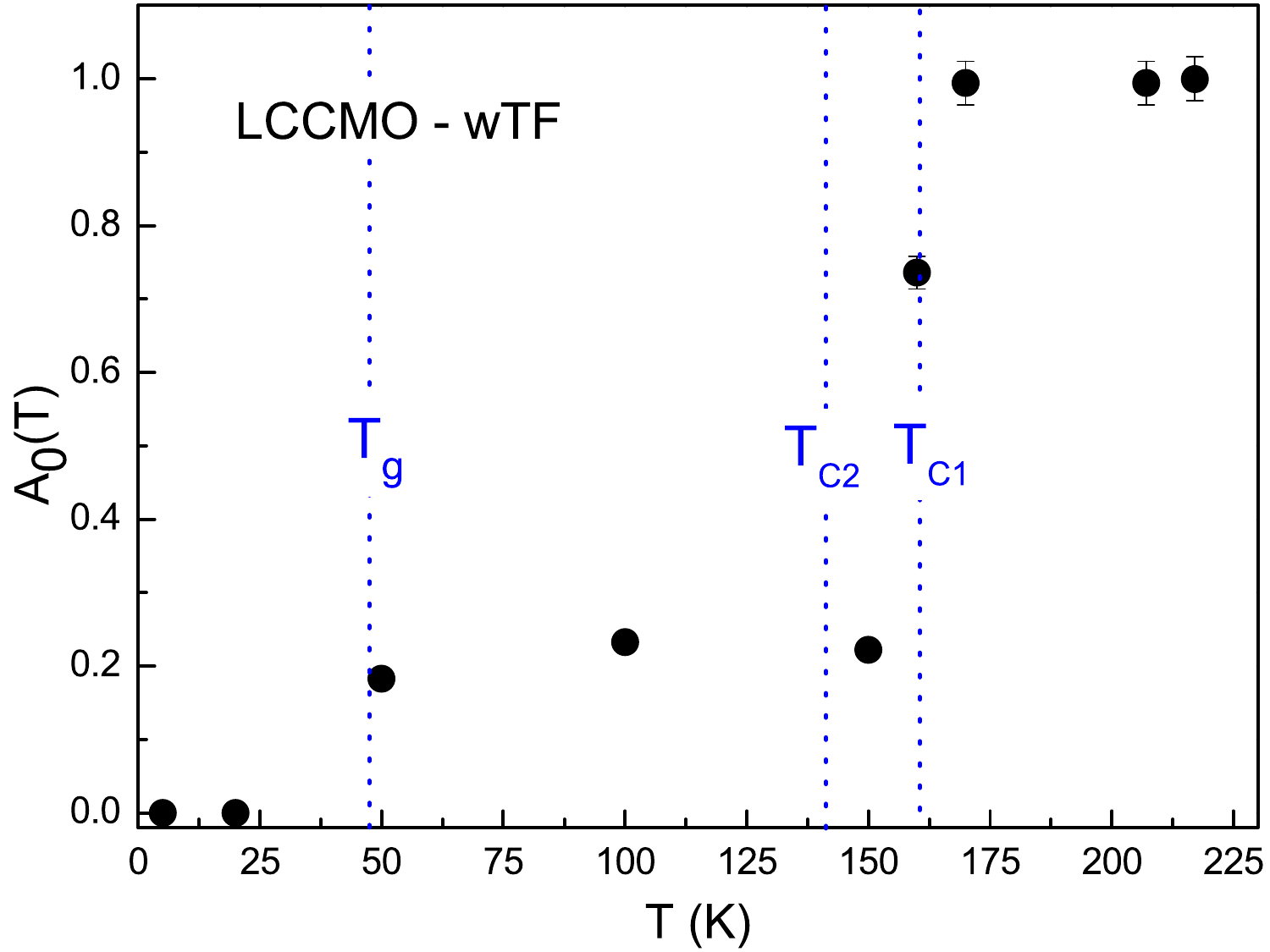}
\end{center}
\caption{Variation of weakly damped paramagnetic fraction with temperature derived from wTF spectra of LSCMO. Indicated marks for $T_{C1}$, $T_{C2}$ and $T_g$ are taken from magnetization and susceptibility data.}
\label{Fig_S_TF_LCCMO}
\end{figure}

Both the ZF and wTF spectra of LCCMO and LSCMO can be interpreted by the same picture conjectured for LBCMO: (i) $T_C$ obtained from the $\mu$SR are in agreement with $M(T)$ results; (ii) below $T_C$ there still exists a PM fraction, \textit{i.e.}, there is a $T$-range with an inhomogeneous magnetic state, with the co-existence of PM, SG and ordered phases.

\subsection{S4: Raman Spectroscopy}

\begin{figure*}
\begin{center}
\includegraphics[width= \textwidth]{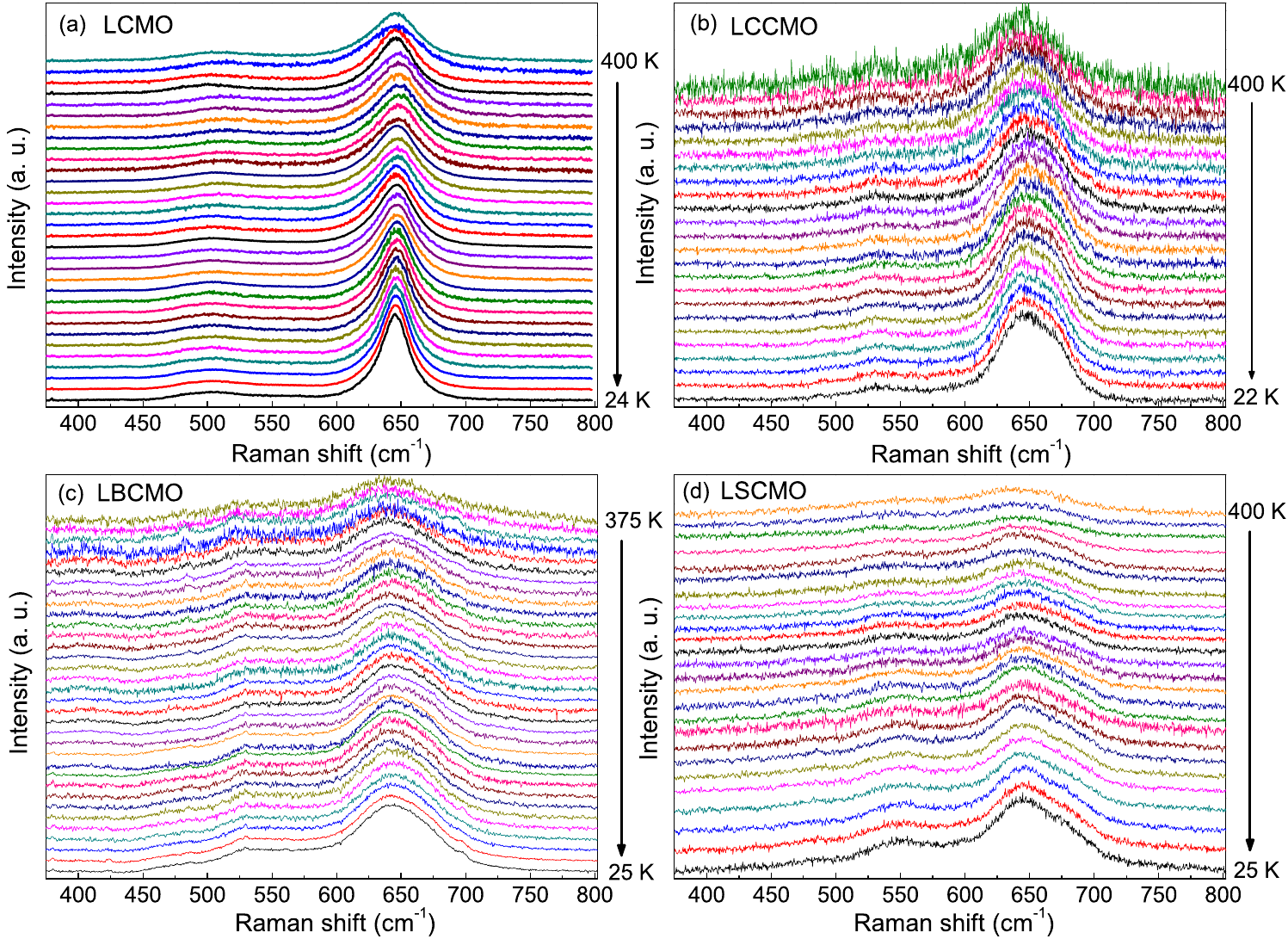}
\end{center}
\caption{Magnified view of the Raman spectra of (a) LCMO, (b) LCCMO, (c) LBCMO and (d) LSCMO at several $T$, highlighting the stretching and anti-stretching/bending modes.}
\label{Fig_Escadinha}
\end{figure*}

Fig. \ref{Fig_Escadinha} shows the unpolarized Raman spectra taken at several $T$ for the La$_{2-x}$A$_{x}$CoMnO$_{6}$ samples here investigated, were the presence of both the stretching and anti-stretching/bending modes is clear.

\begin{figure}[H]
\begin{center}
\includegraphics[width=0.47 \textwidth]{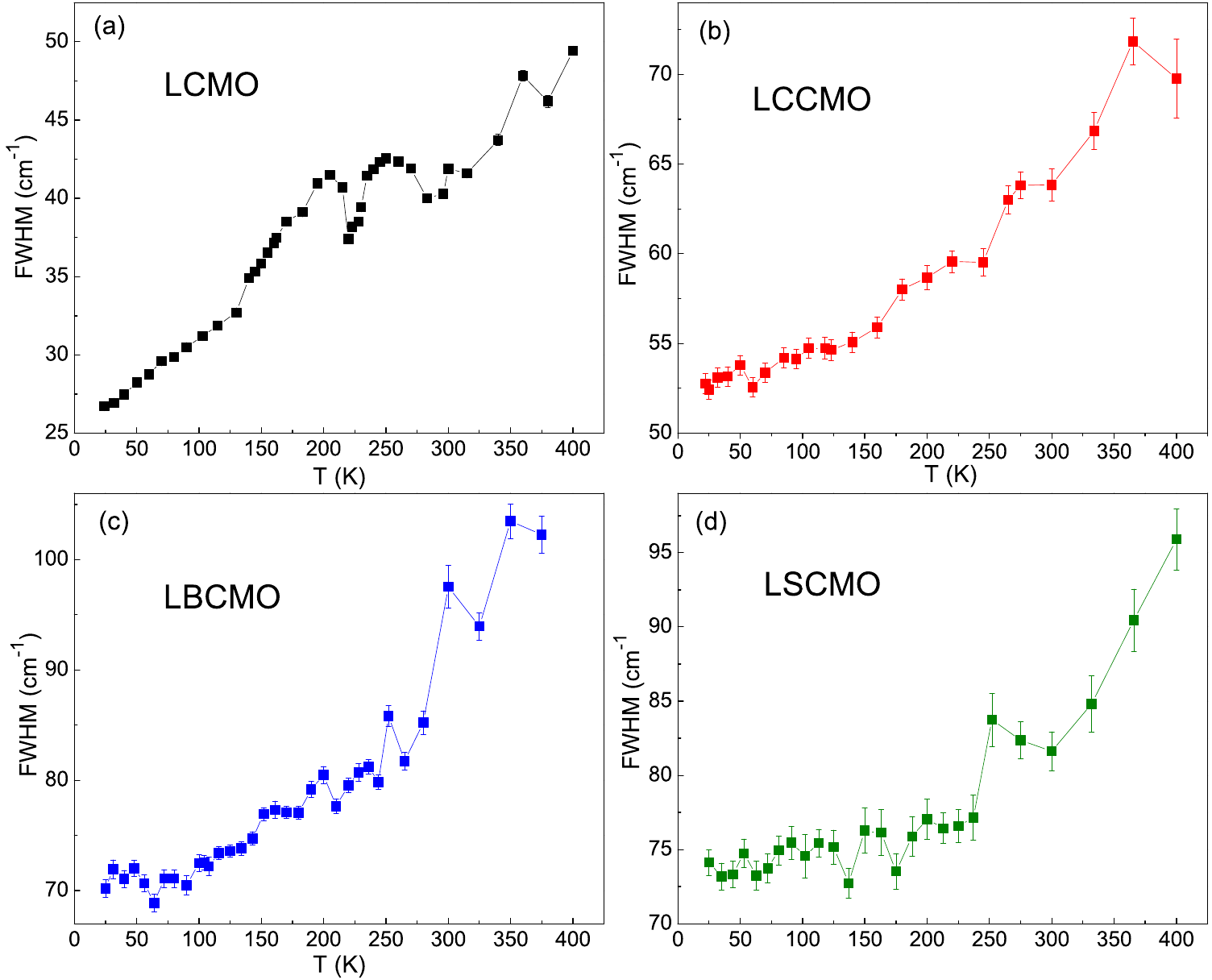}
\end{center}
\caption{FWHM as a function of $T$ for the stretching mode of (a) LCMO, (b) LCCMO, (c) LBCMO and (d) LSCMO. The solid lines are guides for the eye.}
\label{Fig_FWHM_stretching}
\end{figure}

\begin{figure}[H]
\begin{center}
\includegraphics[width=0.47 \textwidth]{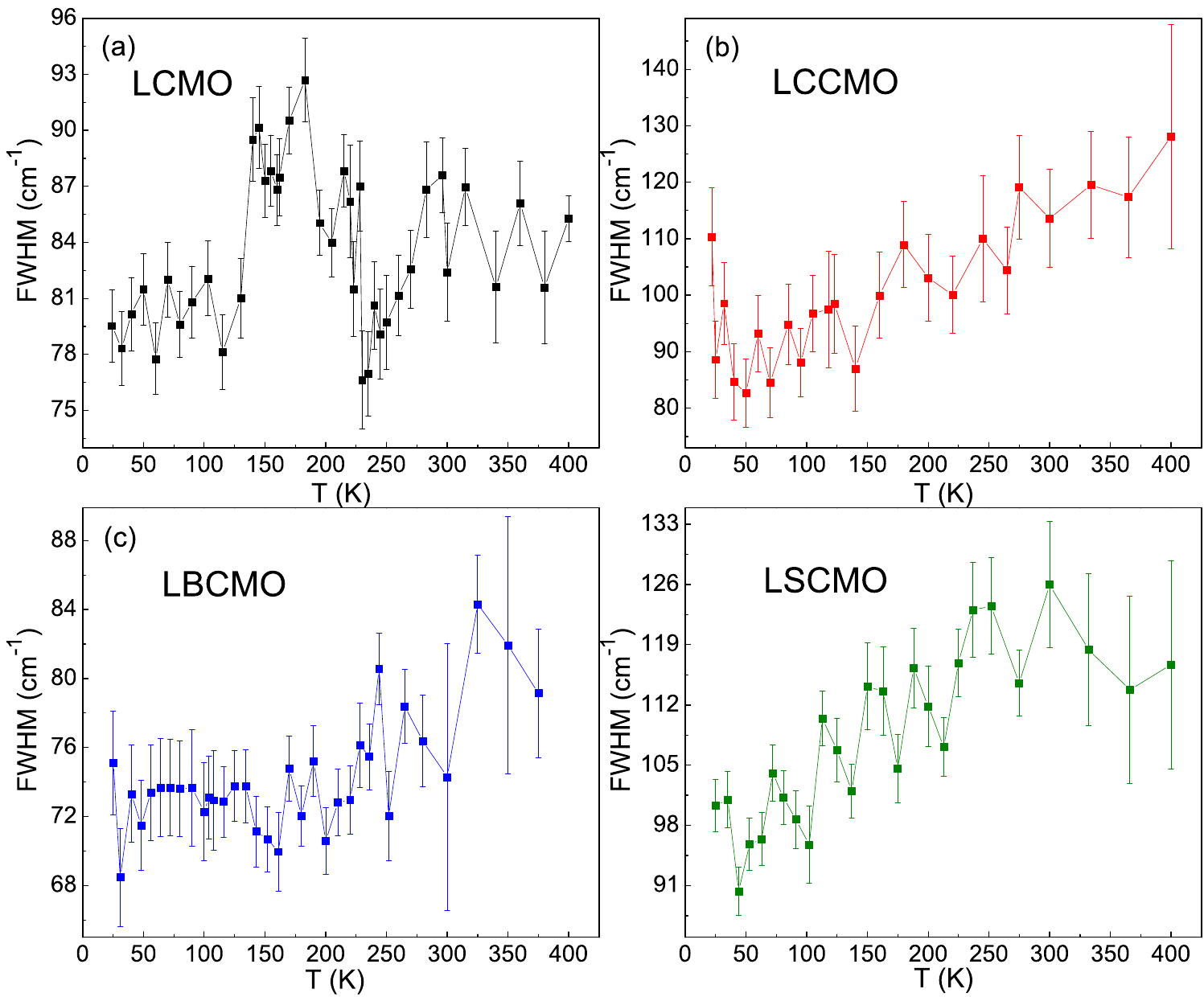}
\end{center}
\caption{FWHM as a function of $T$ in the anti-stretching/bending modes of (a) LCMO, (b) LCCMO, (c) LBCMO and (d) LSCMO. The solid lines are guides for the eye.}
\label{Fig_FWHM_bending}
\end{figure}

\begin{figure}[H]
\begin{center}
\includegraphics[width=0.47 \textwidth]{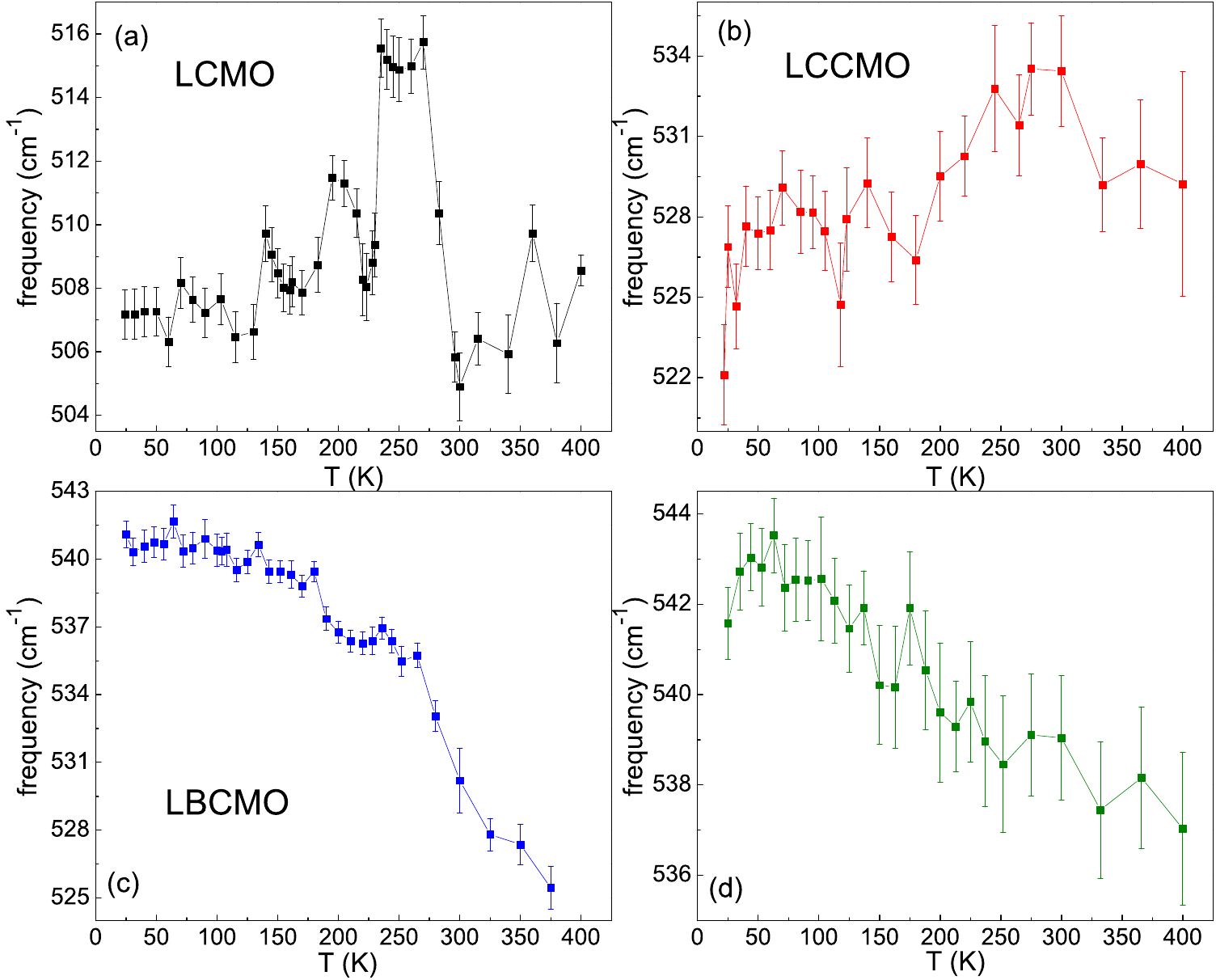}
\end{center}
\caption{Relative shifts as a function of $T$ in the anti-stretching/bending modes of (a) LCMO, (b) LCCMO, (c) LBCMO and (d) LSCMO. The solid lines are guides for the eye.}
\label{Fig_freq_bending}
\end{figure}

Each spectra was fitted with two Lorentzian functions [referencia] in order to obtain the frequency and linewidth (full width at half maxima - FWHM) of both stretching and anti-stretching/bending modes. Figs. \ref{Fig_FWHM_stretching} and \ref{Fig_FWHM_bending} display respectively the FWHM of the stretching and anti-stretching modes, and Fig. \ref{Fig_freq_bending} shows the $T$-dependence of the phonon frequency for the anti-stretching/bending mode.

\subsection{S5: XAS and XMCD}

\begin{figure}
\begin{center}
\includegraphics[width=0.48 \textwidth]{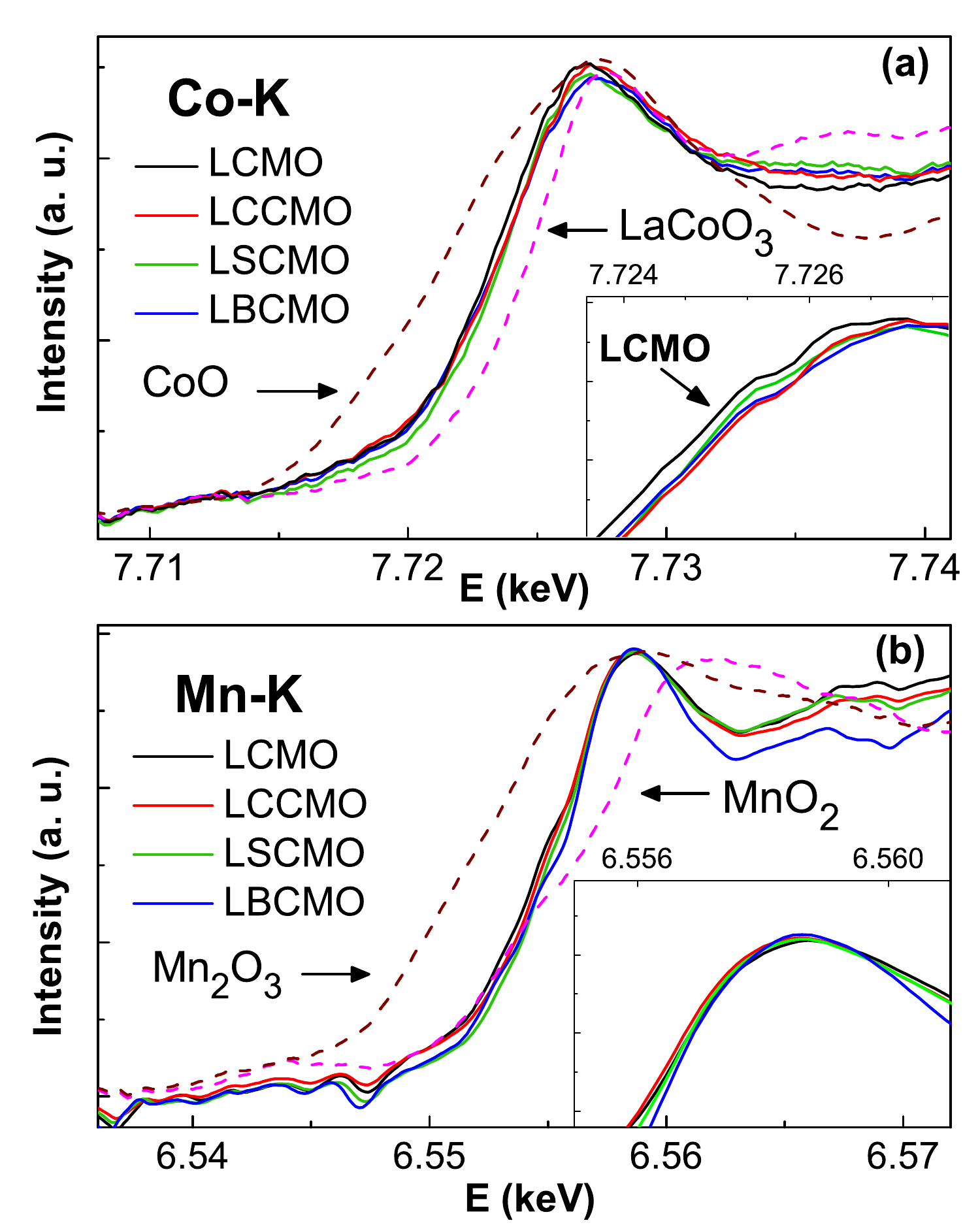}
\end{center}
\caption{Room-$T$ XAS spectra of La$_{2-x}$A$_{x}$CoMnO$_{6}$ at (a) Co-$K$ and (b) Mn-$K$ edges. The spectra of CoO, LaCoO$_{3}$, Mn$_{2}$O$_{3}$ and MnO$_{2}$ are also displayed as reference for Co$^{2+}$, Co$^{3+}$, Mn$^{3+}$ and Mn$^{4+}$ configurations, respectively. The insets show magnified views of the spectra at the absorption regions.}
\label{Fig_XAS_K}
\end{figure}

In order to determine the Co- and Mn-valence states in the La$_{2-x}$A$_{x}$CoMnO$_{6}$ samples, we performed x-ray absorption spectroscopy (XAS) at Co- and Mn-$K$ and $L_{2,3}$ edges. Fig. \ref{Fig_XAS_K}(a) shows the Co-$K$ edge curves of all investigated compounds , for which we used CoO and LaCoO$_3$ as reference samples for Co$^{2+}$ and Co$^{3+}$ configuration, respectively, and Mn$_2$O$_3$ and MnO$_2$ as reference samples for Mn$^{3+}$ and Mn$^{4+}$, respectively. 

The Co valence can be roughly estimated from the comparison of the energy position of the edge jump of each investigated sample to that of the standards for Co$^{2+}$ and Co$^{3+}$ configurations. As can be seen, all curves lie in between those of CoO and LaCoO$_3$, indicating that a fraction of Co$^{3+}$ is present already for the LCMO parent compound. This is in agreement with the results found in the $L_{2,3}$ edges and is corroborated by our SXRD results that show a disordered structural configuration for this sample. Fig. \ref{Fig_XAS_K}(b) shows the Mn-$K$ edge curves. As can be noted, the curves lie in between those of Mn$_2$O$_3$ and MnO$_2$, indicating that Mn is also in mixed valence state in all investigated samples.

\begin{figure}
\begin{center}
\includegraphics[width=0.47 \textwidth]{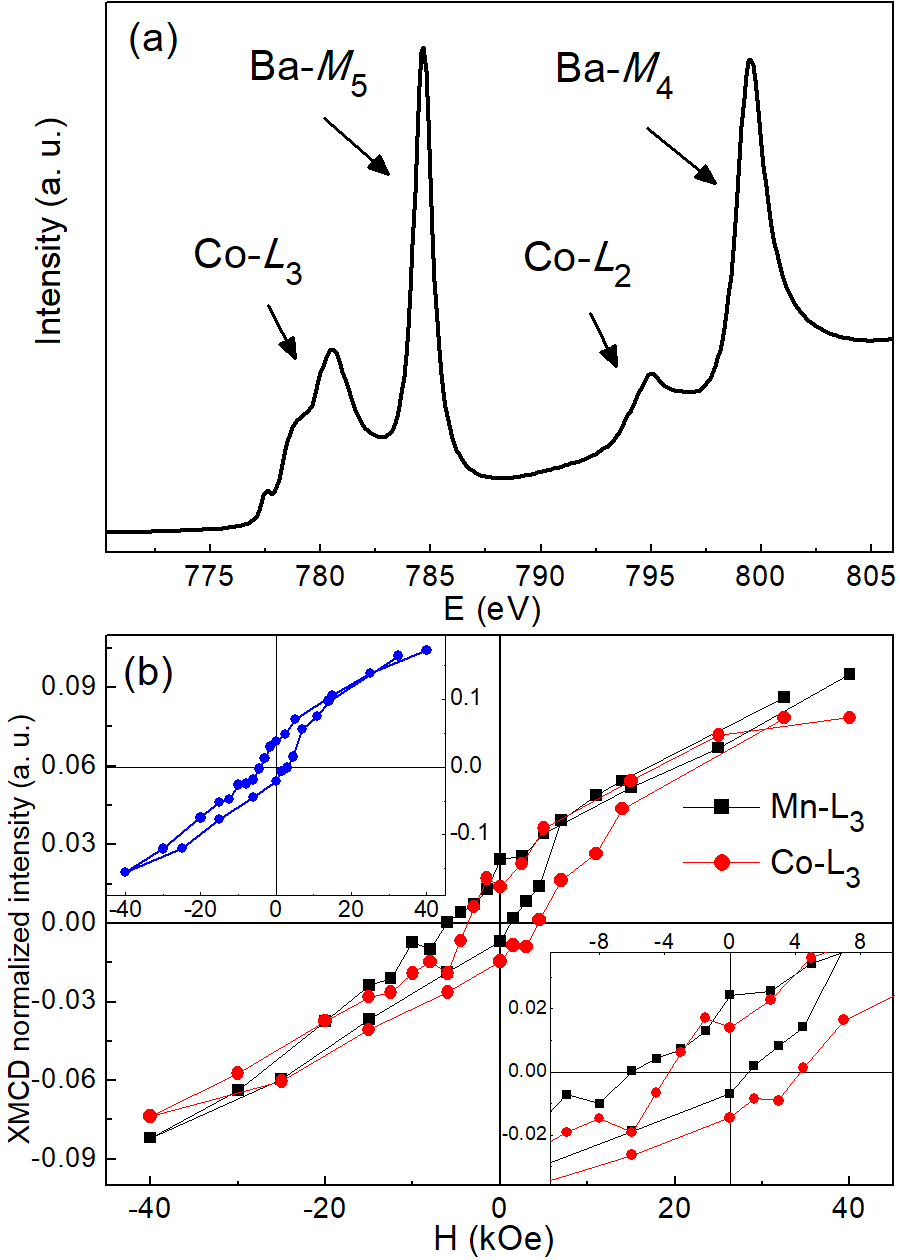}
\end{center}
\caption{(a) Co-$L_{2,3}$ XAS spectra of LBCMO, where the proximity to the Ba-$M_{4,5}$ edges is evident. (b) XMCD hysteresis loops of Co and Mn $L_3$ edges for LBCMO, measured at 14 K. The solid lines are guides for the eye.}
\label{Fig_BaL3}
\end{figure}

The X-ray absorption Spectroscopy (XAS) results obtained for LBCMO at Co-$L_{2,3}$ edges are not displayed in the main text due to its proximity with Ba-$M_{4,5}$ absorption edges. This can be clearly observed in Fig. \ref{Fig_BaL3}(a) below. However, it can be noticed that the spectral region encompassing the Co-$L_3$ edge is similar to that found for LCCMO and LSCMO [Fig. 8(a) of the main text], with the increased peak at the high energy side indicating that also for this sample the hole doping at La$^{3+}$ site acts mainly to increase the Co average valence.

We carried XMCD hysteresis loops at Co- and Mn-$L_3$ edges for LBCMO. Fig. \ref{Fig_BaL3}(b) shows the resulting curves, obtained using $H_{max}$ = 40 kOe at $T$ = 14 K, after ZFC the sample. Despite the presence of Ba-$M_5$ near Co-$L_3$ edge, we can see that the loop is shifted to the right in the same way that was found for LCCMO and LSCMO. Moreover, the Mn-$L_3$ loop, for which there is not such a problem of proximity to other absorption edges, is displaced to the left as was also observed for the Ca- and Sr-doped samples. As expected, the loop resulting from the sum of Co and Mn signals is shifted to the left due to the uncompensated displacements of the Co- and Mn-$L_3$ loops. These results reinforce that the overall effect of hole doping at La site is rather similar in all investigated samples.


\begin{thebibliography}{99}

\bibitem{Dagotto} E. Dagotto, Science \textbf{318}, 1076 (2007).

\bibitem{Nogues} J. Nogu\'{e}s and I. K. Schuller, J. Magn. Magn. Mater. \textbf{192}, 203 (1999).

\bibitem{Geshev} A. Harres, M. Mikhov, V. Skumryev, A. M. H. de Andrade, J. E. Schmidt, and J. Geshev, J. Magn. Magn. Mater. \textbf{402}, 76 (2016).

\bibitem{Saha} J. Saha and R. H. Victora, Phys. Rev. B \textbf{76}, 100405(R) (2007).

\bibitem{Giri} S. K. Giri, R. C. Sahoo, P. Dasgupta, A. Poddar, and T. K. Nath, J. Phys. D: Appl. Phys. \textbf{49}, 165002 (2016).

\bibitem{CoIr_PRB} L. T. Coutrim, E. M. Bittar, F. Stavale, F. Garcia, E. Baggio-Saitovitch, M. Abbate, R. J. O. Mossanek, H. P. Martins, D. Tobia, P. G. Pagliuso, and L. Bufai\c{c}al, Phys. Rev. B \textbf{93}, 174406 (2016).

\bibitem{Maity} T. Maity, S. Goswami, D. Bhattacharya, and S. Roy, Phys. Rev. Lett. \textbf{110}, 107201 (2013).

\bibitem{Wang} B. M. Wang, Y. Liu, P. Ren, B. Xia, K. B. Ruan, J. B. Yi, J. Ding, X. G. Li, and L. Wang, Phys. Rev. Lett. \textbf{106}, 077203 (2011).

\bibitem{Nayak} A. K. Nayak, M. Nicklas, S. Chadov, C. Shekhar, Y. Skourski, J. Winterlik, and C. Felser, Phys. Rev. Lett. \textbf{110}, 127204 (2013).

\bibitem{Prieto} A. Migliorini, B. Kuerbanjiang, T. Huminiuc, D. Kepaptsoglou, M. Mu\~{n}oz, J. L. F. Cu\~{n}ado,
J. Camarero, C. Aroca, G. Vallejo-Fern\'{a}ndez, V. K. Lazarov and J. L. Prieto, Nat. Mater. \textbf{17}, 28 (2017).

\bibitem{ZEBmodel} L. T. Coutrim, E. M. Bittar, F. Garcia, and L. Bufai\c{c}al, Phys. Rev. B \textbf{98}, 064426 (2018).

\bibitem{Vasala} S. Vasala and M. Karppinen, Prog. Solid State Chem. \textbf{43}, 1 (2015).

\bibitem{Serrate} D. Serrate, J. M. De Teresa, and M. R. Ibarra, J. Phys.: Condens. Matter \textbf{19}, 023201 (2007).

\bibitem{Huang} S. Huang, L. R. Shi, Z. M. Tian, H. G. Sun, and S. L. Yuan, J. Magn. Magn. Mater. \textbf{394}, 77 (2015).

\bibitem{Xie} L. Xie and H. G. Zhang, Curr. Appl. Phys. \textbf{18}, 261 (2018).

\bibitem{Murthy} J. Krishna Murthy and A. Venimadhav, Appl. Phys. Lett. \textbf{103}, 25410 (2013).

\bibitem{CaCoMn_JMMM} L. Bufai\c{c}al, R. Finkler, L. T. Coutrim, P. G. Pagliuso, C. Grossi, F. Stavale, E. Baggio-Saitovitch, and E. M. Bittar, J. Magn. Magn. Mater. \textbf{433}, 271 (2017).

\bibitem{SM} See Supplemental Material at for details.

\bibitem{Fujiki} S. Fujiki and S. Katsura, Prog. Theor. Phys. \textbf{65}, 4 (1981).

\bibitem{Balanda} M. Balanda, Acta Phys. Pol. A \textbf{124}, 6 (2013).

\bibitem{Frey} M. H. Frey and D. A. Payne, Phys. Rev. B \textbf{54}, 3158 (1996).

\bibitem{XDS} F. A. Lima, M. E. Saleta, R. J. S. Pagliuca, M. A. Eleot\'{e}rio, R. D. Reis, J. Fonseca J\'{u}nior, B. Meyer, E. M. Bittar, N. M. Souza-Neto, and E. Granado, J. Synchrotron Radiat. \textbf{23}, 1538 (2016).

\bibitem{GSAS} A. C. Larson and R. B. Von Dreele, Los Alamos National Laboratory Report No. LAUR 86-748, 2000; B. H. Toby, J. Appl. Crystallogr. \textbf{34}, 210 (2001).

\bibitem{DXAS} J. C. Cezar, N. M. Souza-Neto, C. Piamonteze, E. Tamura, F. Garcia, E. J. Carvalho, R. T. Neueschwander, A. Y. Ramos, H. C. N. Tolentino, A. Caneiro, N. E. Massa, M. J. Martinez-Lope, J. A. Alonso, and J. Itie, J. Synchrotron Radiat. \textbf{17}, 93 (2010).

\bibitem{PGM} J.C. Cezar, P.T. Fonseca, G.L.M.P. Rodigues, A.R.B. de Castro, R.T. Neuenschwander, F. Rodrigues, B.C. Meyer, L.F.S. Ribeiro, A.F.A.G. Moreira, J.R. Piton, M.A. Raulik, M.P. Donadio, R.M. Seraphim, M.A. Barbosa, A. de Siervo, R. Landers, and A.N. de Brito, J. Phys.: Conf. Ser. \textbf{425} (7), 07201 (2013).

\bibitem{Goodenough} J. B. Goodenough, Phys. Review \textbf{100}, 2 (1955).

\bibitem{Singh} M. P. Singh, K. D. Truong, and P. Fournier, Appl. Phys. Lett. \textbf{91}, 042504 (2007).

\bibitem{Chen} Yi Qi Lin and Xiang Ming Chen, J. Am. Ceram. Soc. \textbf{94}[3], 782 (2011).

\bibitem{Blasco} A. J. Bar\'{o}n-Gonz\'{a}lez, C. Frontera, J. L. Garc\'{i}a-Mu\~{n}oz, B. Rivas-Murias and J. Blasco, J. Phys.: Condens. Matter \textbf{23}, 496003 (2011).

\bibitem{Vashook} V. Vashook, D. Franke, J. Zosel, L. Vasylechko, M. Schmidt, and U. Guth, J. Alloys Compd. \textbf{487}, 577 (2009).

\bibitem{Taraphder} A. Khan, P.R. Mandal, S. Chatterjee, T.K. Nath and A. Taraphder, arXiv:1807.05019v1 (2018).

\bibitem{Xu} L. Xing, Q. Li, and M. Xu, J. Alloys Compd. \textbf{774}, 646 (2019).

\bibitem{Berger} V. A. Cherepanov, E. A. Filonova, V. I. Voronin, and I. F. Berger, J. Solid State Chem. \textbf{153}, 205 (2000).

\bibitem{Shannon} R. D. Shannon, Acta Cryst. \textbf{A32}, 751 (1976).

\bibitem{Dass} R. I. Dass and J. B. Goodenough, Phys. Rev. B \textbf{67}, 014401 (2003).

\bibitem{Burnus} T. Burnus, Z. Hu, H. H. Hsieh, V. L. J. Joly, P. A. Joy, M. W. Haverkort, Hua Wu, A. Tanaka, H.-J. Lin, C. T. Chen, and L. H. Tjeng, Phys. Rev. B \textbf{77}, 125124 (2008).

\bibitem{Joy} P. A. Joy, Y. B. Khollam, and S. K. Date, Phys. Rev. B \textbf{62}, 13 (2000).

\bibitem{Fournier} K. D. Truong, J. Laverdi\`{e}re, M. P. Singh, S. Jandl, and P. Fournier, Phys. Rev. B \textbf{76}, 132413 (2007).

\bibitem{Woodward} C. J. Howard, B. J. Kennedy and P. M. Woodwardd    , Acta Crystaloogr. B \textbf{59},463 (2003).

\bibitem{GKA} J. B. Goodenough, \textit{Magnetism and Chemical Bond} (Interscience, New York, 1963).

\bibitem{Mydosh} J. A. Mydosh, \textit{Spin Glasses: An Experimental Introduction} (Taylor \& Francis, London, 1993).

\bibitem{Souletie} J. Souletie and J.L. Tholence, Phys. Rev. B \textbf{32}, 516(R) (1985).

\bibitem{Malinowski} A. Malinowski, V. L. Bezusyy, R. Minikayev, P. Dziawa, Y. Syryanyy, and M. Sawicki, Phys. Rev. B \textbf{84}, 024409 (2011).

\bibitem{Murthy2} J. Krishna Murthy, K. D. Chandrasekhar, H. C. Wu, H. D. Yang, J. Y. Lin, and A. Venimadhav, J. Phys.: Condens. Matter \textbf{28}, 086003 (2016).

\bibitem{Anand2} V. K. Anand, D. T. Adroja, and A. D. Hillier, Phys. Rev. B \textbf{85}, 014418 (2012).

\bibitem{Sikora} M. Sikora, Cz. Kapusta, M. Borowiec, C. J. Oates, V. Prochazka, D. Rybicki, D. Zajac, J. M. De Teresa, C.
Marquina, and M. R. Ibarra, Appl. Phys. Lett. \textbf{89}, 062509 (2006).

\bibitem{Haskel} C. A. Escanhoela, G. Fabbris, F. Sun, C. Park, J. Gopalakrishnan, K. Ramesha, E. Granado, N. M. Souza-Neto, M. van Veenendaal, and D. Haskel, Phys. Rev. B \textbf{98}, 054402 (2018). 

\bibitem{Iliev} M. N. Iliev, M. V. Abrashev, A. P. Litvinchuk, V. G. Hadjiev, H. Guo, and A. Gupta, Phys. Rev. B \textbf{75}, 104118 (2007). 

\bibitem{Murthy3} J. Krishna Murthy, K. D. Chandrasekhar, S. Murugavel, and A. Venimadhav, J. Mater. Chem. C \textbf{3},836 (2015).

\bibitem{Silva} R. X. Silva, H. Reichlova, X. Marti, R. Paniago, and C. W. A. Paschoal, arXiv:1802.07326v2 (2018).

\bibitem{Cardona} M. Cardona, \textit{Light Scattering in Solids I} (Springer, Berlin, 1983).

\bibitem{Granado} E. Granado, A. Garc\'{i}a, J. A. Sanjurjo, C. Rettori, I. Torriani, F. Prado, R. D. S\'{a}nchez, A. Caneiro, and S. B. Oseroff, Phys. Rev. B \textbf{60}, 11879 (1999).

\bibitem{Gervais} F. Gervais and B. Piriou, Phys. Rev. B \textbf{11}, 10 (1975).

\bibitem{Mir} L. L\'{o}pez-Mir, R. Galceran, J. Herrero-Mart\'{i}n, B. Bozzo, J. Cisneros-Fern\'{a}ndez, E. V. Pannunzio Miner, A. Pomar, L. Balcells, B. Mart\'{i}nez, and C. Frontera, Phys. Rev. B \textbf{95}, 224434(2017).

\bibitem{Thole} B. T. Thole, P. Carra, F. Sette, and G. van der Laan, Phys. Rev. Lett. \textbf{68}, 1943 (1992).

\bibitem{Carra} P. Carra, B. T. Thole, M. Altarelli, and X. Wang, Phys. Rev. Lett. \textbf{70}, 694 (1993).

\bibitem{Teramura} Y. Teramura, A. Tanaka, and T. Jo, J. Phys. Soc. Jpn. \textbf{65}, 4 (1996).

\bibitem{Groot} C. Piamonteze, P. Miedema, and F. M. F. de Groot, Phys. Rev. B \textbf{80}, 184410 (2009).

\bibitem{Hollmann} N. Hollmann, M. W. Haverkort, M. Benomar, M. Cwik, M. Braden,1 and T. Lorenz, Phys. Rev. B \textbf{83}, 174435 (2011).

\bibitem{Miao} X. Miao, L. Wu, Y. Lin, X. Yuan, J. Zhao, W. Yan, S. Zhou, and L. Shi, Chem. Commun. \textbf{55}, 1442 (2019).

\end{thebibliography}
\end{document}